\documentclass[10pt,journal,compsoc]{IEEEtran}

\usepackage{cite}
\usepackage{amsmath,amssymb,amsfonts}
\usepackage{amsmath}
\usepackage{algorithmic}
\usepackage{graphicx}
\usepackage{textcomp}
\usepackage{xcolor}
\usepackage{todonotes}
\usepackage{amsmath}
\usepackage{caption}
\usepackage{subcaption}
\usepackage{hyperref}
\usepackage{booktabs}
\usepackage{amssymb}
\usepackage{pifont}
\usepackage{multirow}
\usepackage{pdfpages}
\usepackage{tikzpagenodes} 
\usepackage{cite}
\usepackage{booktabs}
\usepackage{cleveref}

\DeclareMathOperator*{\argmax}{arg\,max}

\begin{document}








\title{Dynamics of Collective Group Affect: Group-level Annotations and the Multimodal Modeling of Convergence and Divergence}






\author{Navin~Raj Prabhu, Maria~Tsfasman, Catharine~Oertel, Timo~Gerkmann, and~Nale~Lehmann-Willenbrock

\IEEEcompsocitemizethanks{\IEEEcompsocthanksitem Navin Raj Prabhu and Timo Gerkmann are with the Signal Processing Lab, University of Hamburg, Germany, 20146. \protect E-mail: navin.raj.prabhu@uni-hamburg.de, timo.gerkmann@uni-hamburg.de

\IEEEcompsocthanksitem Maria Tsfasman and Catharine Oertel are with the Department of Intelligent Systems, TU Delft, 2628 Delft, The Netherlands. \protect E-mail: m.tsfasman@tudelft.nl, c.r.m.m.oertel@tudelft.nl

\IEEEcompsocthanksitem Nale Lehmann-Willenbrock is with the Department of Industrial and Organizational Psychology,  University of Hamburg, Germany, 20146.\protect\\ E-mail: nale.lehmann-willenbrock@uni-hamburg.de
}%

\thanks{This work was funded under the Excellence Strategy of the Federal Government and the Länder, and the project ''Mechanisms of Change in Dynamic Social Interaction'' (LFF-FV79, Landesforschungsförderung Hamburg).}
}

\markboth{Journal of \LaTeX\ Class Files,~Vol.~xx, No.~xx, xxx 2022}%
{Raj Prabhu \MakeLowercase{\textit{et al.}}: Dynamic Group Affect: Annotation Procedure, Quantitative Analysis and Modeling}

\IEEEtitleabstractindextext{%
\begin{abstract}
Collaborating in a group, whether face-to-face or virtually, involves continuously expressing emotions and interpreting those of other group members. Therefore, understanding group affect is essential to comprehending how groups interact and succeed in collaborative efforts. In this study, we move beyond individual-level affect and investigate group-level affect---a collective phenomenon that reflects the shared mood or emotions among group members at a particular moment. As the first in literature, we gather annotations for group-level affective expressions using a fine-grained temporal approach (15 second windows) that also captures the inherent dynamics of the collective construct. To this end, we use trained annotators and an annotation procedure specifically tuned to capture the entire scope of the group interaction. In addition, we model group affect dynamics over time. One way to study the ebb and flow of group affect in group interactions is to model the underlying convergence (driven by emotional contagion) and divergence (resulting from emotional reactivity) of affective expressions amongst group members. To capture these interpersonal dynamics, we extract synchrony based features from both audio and visual social signal cues. An analysis of these features reveals that interacting groups tend to diverge in terms of their social signals along neutral levels of group affect, and converge along extreme levels of affect expression. We further present results on the predictive modeling of dynamic group affect which underscores the importance of using synchrony-based features in the modeling process, as well as the multimodal nature of group affect. We anticipate that the presented models will serve as the baselines of future research on the automatic recognition of dynamic group affect. 
\end{abstract}

\begin{IEEEkeywords}
Group affect, affect dynamics, annotations, convergence, divergence, multimodal analysis, automatic affect recognition 
\end{IEEEkeywords}

}

\maketitle

\IEEEdisplaynontitleabstractindextext
\IEEEpeerreviewmaketitle

\ifCLASSOPTIONcompsoc
\IEEEraisesectionheading{\section{Introduction}\label{sec:introduction}}
\else
\section{Introduction}\label{sec:introduction}
\fi

Group affect is a collective social construct that represents the jointly experienced shared mood or emotions that group members hold in common at a given point in time \cite{knight2014gapositivenegative}. There is scholarly consensus that group affect is characterized by both (i) \textit{bottom-up} and (ii) \textit{top-down} affective processes \cite{barsade2015group}. As a \textit{bottom-up} group process, group affect manifests as the “sum of its parts”, i.e., the collective aggregate of individual group members’ affective states and traits. As a \textit{top-down} group process, group affect is viewed "as a whole", with characteristics and properties of the group acting upon the emotions of the individuals within it. Both types of processes require capturing group affect as a dynamic and interactive social phenomenon \cite{jones2021observational}. 
A growing literature highlights the relevance and consequences of collective affective constructs, specifically regarding how group affect relates to group processes and outcomes \cite{kozlowski2006enhancing, collins2013group, lei2015affect, dhall2015more, jones2021observational, veltmeijer2021automatic, sharma2021audio}. For example, a widely cited review of the literature discusses the importance of group affect for shaping (i) group member attitudes, (ii) cooperation and conflict resolution, (iii) group creativity and decision making, and (iv) group effectiveness and performance \cite{barsade2015group}. However, there is a notable dearth of empirical and quantitative research on group affect as it emerges and develops during group interactions \cite{barsade2015group, jones2021observational}. 
From a review of the literature on group affect, spanning both organizational psychology and computer science, we see five key shortcomings and associated challenges.

First, whereas extensive research has focused on affect as an individual-level construct \cite{Schuller2018-xi, kusha_valence, prabhuUncertTAFFC24}, we know much less about group affect as a collective interactional phenomenon (for an overview, see \cite{veltmeijer2021automatic}). It is important to acknowledge that the annotation and analysis of \textit{collective group affect} and the related social signals is considerably more complex compared to the analysis of individual members' affect \cite{dhall2015more, sharma2021audio}. Moreover, capturing the collective and shared affect of a group requires considering all the group members and their interactions as a whole, instead of simply averaging individual affect expressions. 


Second, the scarcity of prior insights into group affect may also be due to the dynamic and often "messy" nature of group interactions, which makes the phenomenon of collective group affect subject to \textit{temporal dynamics and change}. Here, we consider "group interactions" to be any task-directed collaborative social interactions between two or more group members (e.g., group discussions and team meetings). 
The limited literature on group affect to date has predominantly studied it by collecting annotations of static images \cite{dhall2015more, dhall2017GAchallenge, wang2023implementing} or temporally independent video segments \cite{huang2018multimodal, ghosh2018automatic, sharma2021audio} \textit{without accounting for any temporal context} \cite{veltmeijer2021automatic}. There is a feasibility argument to be made here in both these cases. Static images are much easier to obtain compared to observational group interaction data. Similarly, annotating video segments in an temporally independent manner without the temporal context is resource efficient, where annotators need not continuously track the group's affective expressions and its associated changes between consecutive segments. However, such an annotation approach is agnostic to the inherently {dynamic} nature of group affect \cite{barsade2015group, lei2015affect, veltmeijer2021automatic}, which limits the resulting empirical contributions and represents a misalignment between the theoretical construct of group affect as a dynamic process and its measurement \cite{jones2021observational, nale_dyninterpersonal2025}. 

Third, capturing collective group affect and its temporal dynamics is not feasible without training annotators. Any annotation process that accounts for the inherent temporal dynamics and change in group affect (e.g., thin-sliced annotations \cite{ambady1992thin}) is time consuming because of the scope of the entire group interaction that has to be attentively tracked by annotators \cite{veltmeijer2021automatic}. Perceiving the group interaction involves simultaneously observing all group members and their affective expressions displayed across modalities (i.e., nonverbal, paraverbal, as well as verbal cues that convey affective content). Such annotations of complex group-level constructs usually require extensively trained annotators \cite{raj2020defining, ramanrajprabhu2023}. However, investing such efforts is necessary to advance our empirical understanding of group affect. 

Fourth, in a summary of the organizational psychology literature on group affect, Barsade \cite{barsade2015group} identified a crucial research need regarding process-oriented research and corresponding \textit{fine-grained analyses} of the temporal dynamics of group affect. In particular, the study of \textit{convergence} and \textit{divergence} in affective expressions are ways to study the ebb and flow of group affect. For example, Hareli and Rafaeli \cite{hareli2008emotion} proposed a cyclical theoretical model where individual emotional expressions over time either \textit{converge} through emotion contagion or \textit{diverge} through reactivity (constituting for bottom-up phenomena), and further results in a feedback loop that in turn influences group functioning (top-down phenomena), thereby resulting in the ebb and flow of group affect. Despite the intuitive appeal of this theorizing however, empirical insights into affective convergence and divergence are lacking to date. 

Fifth, from a computer science perspective, the collection of annotations for dynamic group affect allows for the \textit{development of social signal processing techniques} that aptly model the dynamics underlying group affect. The development of such techniques can help to achieve three crucial research gaps in psychology literature: (1) the automatic recognition of group affect with more real-world application possibilities, coping with changing affect over time, and higher robustness \cite{veltmeijer2021automatic}, (2) the fine-grained analysis of affect dynamics \cite{barsade2015group, kozlowski2015advancing}, and (3) the incorporation of multimodal social signals to model group affect \cite{waller2018systematic, nalecohesion2024}. Social signals (e.g., facial gestures, vocal pitch) carry micro-level behavioral information (e.g., synchrony and convergence \cite{delaherche2012interpersonal}) that explain the emergence of dynamic interpersonal relationships \cite{delaherche2012interpersonal} and of group-level social constructs \cite{Nanninga2017-bi, ramanrajprabhu2023}. Similarly, as annotators perceive the group interaction in a multimodal manner when asked to annotate the group as a whole, to aptly study group affect, modeling techniques should incorporate social signals from multiple modalities.

To overcome these limitations, this work takes an important first step by collecting group affect annotations using trained annotators that account for the group as a whole while also capturing the inherent temporal dynamics of collective group affect. We code these group affect annotations using the multimodal longitudinal meeting corpus (MeMo) \cite{memo} that consists of videos of online group interactions involving a group discussion task. The rich audiovisual data and the extensive group discussions in this corpus allow us to study collective group affect with the temporal context and its dynamics taken in to account. Building on the collected annotations for dynamic group affect, we extract multimodal dynamic interpersonal relationship-based features (i.e., audio-visual synchrony and convergence-based features \cite{delaherche2012interpersonal, ramanrajprabhu2023}) to study the affective convergence and divergence among team members, which represents substantial empirical progress in line with conceptual arguments from the literature \cite{barsade2015group}. Furthermore, using the multimodal dynamic interpersonal relationship-based features we also perform predictive modeling of dynamic group affect. These results will serve as the baseline for future research on the automatic recognition of dynamic group affect, using the novel annotations collected as part of this work.

The remainder of this paper is organized as follows. In Section~\ref{sec:background}, we review the extant literature on group affect across disciplines, both from a psychology and from a computer science perspective. In Section~\ref{sec:GAannotations}, we discuss the conceptualization of group affect as actual observable group behavior and the annotation procedure used to aptly capture the dynamics of group affect. In Section~\ref{sec:modeling}, we present the experimental setup and the results on the modeling of dynamic group affect. This includes an analysis of affective convergence and divergence (in Sec.~\ref{subsec:convergedivergence}), as well as results on the automatic prediction of dynamic group affect (in Sec.~\ref{subsec:predmodel}) from social signals extracted during the observed group interactions. We conclude in Section~\ref{subsec:conclusion} with a discussion of our findings, remaining limitations, and opportunities for future research. 


\section{Background and Related Work}\label{sec:background}
In social interactions, both spontaneous (e.g., cocktail parties \cite{ramanrajprabhu2023}) and task-directed (e.g., group discussions \cite{memo}) low-level interaction patterns and individual-level social behaviour result in the emergence of collective \textit{group-level} social constructs. Group affect is one such phenomenon that requires investigations of dynamic social interactions and patterns of interpersonal behavior in order to understand how and why it emerges (for an overview, see \cite{jones2021observational, lehmann2018modeling}).

\subsection{Affect in interacting groups}

Research on affect has focused both on individual-level \cite{Schuller2018-xi, prabhuUncertTAFFC24} and group-level affect \cite{lei2015affect, dhall2017GAchallenge, mou2019alone}. Whereas individual affect has become a well researched topic in the past few decades \cite{Schuller2018-xi}, group-level affect has received comparatively less research attention \cite{barsade2015group}. Empirical research \cite{smith2007can} has demonstrated that group-level emotions are distinct from individual-level affect and that they are shared amongst interlocutors of the interaction. Furthermore, previous research points to crucial consequences of emerging group affect in task-directed social interactions, showing that group affect shapes the group's performance, conflict, creativity, and decision making (e.g., \cite{barsade2015group, hareli2008emotion}). As such, understanding group affect is a key element of understanding how groups interact and achieve collaborative performance \cite{barsade2015group, jones2021observational}.


\subsection{Defining and Quantifying group affect}

Barsade and Knight define group affect as the amalgamation of group members’ affective states (i.e., bottom-up processes) and the mutual influence of a group’s affective context (i.e., top-down processes) \cite{barsade2015group}. More recently, theorists have expanded on the temporal dynamics of group affect, showing how momentary individual- and group- affective experiences become inputs for future group affective experiences \cite{walter2008positive, hareli2008emotion}. It has also been noted that only a few research works have empirically examined this dynamic interplay of group affect and group member experiences \cite{barsade2015group}. We anticipate that the annotations and preliminary quantitative analyses presented in this work will enable such research in the future.

To analyse affect in groups, quantification of affect is the first step. Two types of quantification techniques have been widely used in literature: (1) a categorical affect model using Ekman's six basic emotions \cite{ekman1992there} (e.g., happy, angry, sad) or (2) Russell's circumplex model \cite{russell1980circumplex} where affect is quantified using two continuous, bipolar, and orthogonal dimensions of arousal and valence. Lately, research works and state-of-the-art datasets in affect recognition have moved on from categorical representation to the circumplex model, owing to the fact that the circumplex model is more suitable for capturing the ambiguous and fuzzy nature of affective expressions \cite{sethu2019ambiguous, lei2015affect, prabhuUncertTAFFC24}. 



\subsection{Dynamics inherent in group affect}

Theorizing on group affect as a dynamic, temporally evolving social construct characterised by {bottom-up} and {top-down} processes, while intuitively appealing, has received only limited empirical validation efforts in the psychology literature. This is likely due to the complexity of the phenomenon and the challenge of modeling collective group expressions while also accounting for temporal dynamics. 
Similarly, research on the automatic recognition of group affect \cite{dhall2015more, dhall2017GAchallenge, ghosh2018automatic, wang2023implementing} has not been able to account for the theoretical conceptualization of dynamic group affect. Existing research highlights three crucial challenges in  collecting annotations for dynamic group-level constructs such as group affect, namely (i) time and cost inefficiency \cite{wang2023implementing}, (ii) the difficult nature of the construct \cite{lei2015affect, sharma2021audio}, and (iii) training of annotators as a prerequisite \cite{raj2020defining, ramanrajprabhu2023}.

The appropriate window size required to capture the dynamics of emerging group-level social constructs remains an open research question. For example, Mo et al. \cite{mou2019alone} used 20\,secs windows, whereas Lei et al. \cite{lei2015affect} used 2\,mins windows to capture group affect. Furthermore, we note that recent research on the automatic recognition of \textit{individual}-level affect has moved from segment level annotations ($\approx$ 5\,secs) \cite{MspPod} to much smaller time windows, for example as small as 10\,ms \cite{MspConv, recolaDB}. 
Annotating \textit{larger} time windows is a much simpler task where annotators need not track micro-level behaviours and events. This makes the annotation procedure in general more feasible and resource efficient \cite{wang2023implementing}. However, larger time windows are less appropriate for capturing the dynamics of group affect in dynamic group interactions. In contrast, annotating \textit{smaller} time windows is much more complex and time consuming but better captures the dynamics of group affect  \cite{MspConv}. 


To address the above challenges and aptly capture the dynamic fluctuations of group affect, we use a window size that is iteratively tuned with respect to the construct and the social setting at hand (see Sec.~\ref{sec:tuning-annot-procedure}). 
The annotation strategy, using the temporal context, also takes in to account the top-down and bottom-up processes that characterize the emergence of group-level affect. 


\subsection{Fine-grained analyses of dynamic group affect}
Research on affect modeling at the \textit{individual} level has most frequently used social signals such as video-based facial cues \cite{tzirakis2021-mm, baltruvsaitis2016openface} and audio-based prosodic cues \cite{Schuller2018-xi, prabhuUncertTAFFC24}. Recent research works have revealed the multimodal nature of affect, such that individuals express their affective state across modalities. For example, the audio modality has shown to be more informative of the arousal dimension of affect, whereas the video and text modalities better explain the valence dimension \cite{tzirakis2021-mm, wagner2023dawn, deoliveira23_interspeech, kusha_valence}. As audio features, the pitch, speech rate, intensity and mel-frequency cepstral coefficients (MFCCs) are commonly used \cite{eyben2015geneva}. More recently, embeddings extracted from pretrained transformer and convolution based neural networks have been used (e.g., VGGish \cite{hershey2017cnn}, {wav2vec} \cite{wagner2023dawn}, HuBERT \cite{deoliveira23_interspeech}). As video features individual-level facial action units (AUs) and face pose are commonly used \cite{tian2001recognizing, baltruvsaitis2016openface}. Moreover, embeddings from convolution-based pretrained networks (e.g., ResNet50 \cite{he2016deep}) have also been used to predict individual-level affect.


Moving beyond the analysis of individual-level affect, investigations of collective group-level affect allows insights into affective convergence and divergence processes \cite{hareli2008emotion}. Previous works \cite{barsade2018emotional, hareli2008emotion} highlighted emotional contagion and reactivity as key explanatory mechanisms of how convergence and divergence evolves withing social groups. Emotional contagion and reactivity are processes in which a person or group influences the affect or behavior of another person or group through the conscious or unconscious induction of emotion states and behavioral attitudes \cite{hareli2008emotion}. As a prerequisite for emotional contagion, social interactants need to express their emotional states via social interaction behaviors, to which others can respond by showing similar behavior (i.e., convergence) or dissimilar behavior (i.e., divergence). Capturing this dynamic and reciprocal induction of micro-level behaviors among group members requires the extraction of \textit{synchrony} and \textit{convergence} features, as previously presented in \cite{delaherche2012interpersonal} for dyads within groups. For example, group members may express more activated speech adapting to other group members' level of activation (i.e., emotional contagion), or they might gasp in response to a group member's ill-fated story (emotion reactivity). In this work, we extract these interpersonal relationship-based features in order to account for micro-level emotional and behavioural contagion and reactivity between dyads, to perform fine-grained analyses on the convergence and divergence processes underlying dynamic group affect.

Group-level descriptors are required to study collective group constructs \cite{Nanninga2017-bi}. Existing research has employed aggregation techniques to extract {group}-level features from the above described individual- and dyadic-level features. For example, calculating the \texttt{average}, \texttt{standard-deviation}, \texttt{gradient}, \texttt{minimum} and \texttt{maximum} \cite{Nanninga2017-bi, ramanrajprabhu2023}. The idea here is that the these aggregations describe the distribution of individual- and dyadic-level features within a group. The \texttt{average} and \texttt{standard-deviation} explains the average and the deviation from the average of synchrony measures within the possible dyads in a group. Similarly, the \texttt{minimum} and \texttt{maximum} aggregators represent the least and most synchronous dyad. The \texttt{gradient} aggregator explains the deviation between the least and most synchronous dyad, i.e., the absolute difference between the \texttt{minimum} and \texttt{maximum}.

\subsection{Automatic recognition of dynamic group affect}

Dhall et al. \cite{dhall2015more} pioneered research on the automatic recognition of group affect using static images sourced from the web. The images, consisting of groups of people, were annotated for six categories of happiness intensities. Sharma et al. \cite{sharma2021audio}, Ghosh et al. \cite{ghosh2018automatic}, and Huang et al. \cite{huang2018multimodal} extended this into a multimodal setting with videos, and annotated for three categories of valence (i.e., negative, neutral, and, positive). Building on these works, \cite{dhall2018emotiw, dhall2017GAchallenge} setup challenges for the automatic recognition of group affect. Despite the advances made, these works had two crucial drawbacks: (1) did not account for the temporal dynamics and change in group affect between consecutive segments  (i.e., annotations on temporally independent video segments), and, (2) group affect was not operationalized in a group interactions setting.

Lei et al. \cite{lei2015affect} attempted to solve these drawbacks by annotating 2\,mins of interaction segments using the temporal context in group interaction videos. However, the large window size of 2\,mins does not allow for a fine-grained analysis on the dynamics of group affect. Mou et al. \cite{mou2019alone} made progress by using a smaller window size of 20\,secs. However, the study was not operationalized in an interactions setting, but in a much more simpler setting of a group of participants watching videos and expressing affect in terms of facial and bodily gestures. Wang et al. \cite{wang2023implementing}, more recently, proposed a graph neural network architecture to model group affect using static images. Their contribution was limited by the lack of annotations on the dynamics of group affect, which means that the bottom-up and top-down processes of group affect emergence could not be modeled.

\section{Dynamic Group Affect:\\ Conceptualisation and Annotation}\label{sec:GAannotations}


In line with the literature, we conceptualize group affect as a dynamic and continually evolving social phenomenon in groups (e.g., \cite{barsade2015group, jones2021observational}). Specifically, we define \textit{group affect} as the collective affective state of the group, which is the amalgamation of group members' affective states expressed during group interactions (i.e., a \textit{bottom-up} process) and which in turn affects future affective experiences of the group (i.e., \textit{top-down} influence). Of note, this amalgamation requires individual behaviour to be expressed in terms of social signals which can be annotated by external annotators. 
In this section, we present the annotation strategy employed to collect annotations for dynamic group affect. 
    

\subsection{Dataset}

To investigate dynamic group affect in the context of a social interactions, we used the multimodal longitudinal meeting corpus (MeMo) \cite{memo}, consisting of 45 unscripted video-call discussions in groups of three to six participants. The recorded group interactions lasted for approximately 45\,mins ($\mu=$ 41\,mins and 35\,secs; $\sigma=$ 7\,mins and 30\,secs). As a longitudinal meeting corpus, participants were divided into 15 groups, and each group met 3 times over the course of 2 weeks using a video-conferencing software. At the start, the participants were reported to be complete strangers, having never met each other before the first session. 

The interactions were guided by professional facilitators in order to promote active discussions among the group members. The selected facilitators were experienced in moderating meetings, facilitating creative sessions, and conducting interviews. To achieve maximum affect elicitation the topic of group discussions in MeMo was chosen to be of COVID-19, recorded in the year 2021. Furthermore, to maximise the diversity of in-group opinions, participants were recruited from various COVID-19 affected demographics in every group, e.g., parents with young children, adults of age 50+, students, (ex)-business owners. Overall, 15 groups, totalling 53 participants (25 Male, 28 Female; 18-76 years old) and 4 moderators (3 Male, 1 Female; 24-45 years old) took part in the interactions collected as part of MeMo.



The group affect annotations developed in this work will be released along with the MeMo corpus described in \cite{memo}.

\noindent\paragraph*{\textit{\textbf{Dataset Preprocessing}}} For the purpose of this work, we used only the spontaneous interaction segments from the group interactions within the MeMo corpus. All the interactions in MeMo had eye-gaze calibration and administrative tasks at the beginning and at the end, respectively. Furthermore, in some interactions, participants were late to join the interaction, leading the group to having missing participants. To this, we cropped these segments (i.e., calibration, administrative and missing-participants segments) out of all the interaction videos. The timestamps of these segments were manually annotated by the lead author. 





\subsection{Initial Pilot Studies: Tuning the Annotation Procedure and Annotator Training}\label{sec:tuning-annot-procedure}


We trained annotators in order to address the challenges identified in prior research works \cite{lei2015affect, mou2019alone, sharma2021audio}. We recruited annotators who were organizational psychology students either at the bachelors or the masters level. The recruited annotators were made sure to have had the topic of affect in groups as part of their curriculum and also have prior experiences in annotating for behavioural events in social interactions. In total, we recruited eight annotators (3 Male and 5 Female; 18-25 years old) and made sure at least six annotators annotated a whole interaction recording.


\noindent\paragraph*{\textit{\textbf{Developing the Annotation Procedure and Training the Annotators}}}
We began with an initial pilot study for annotating dynamic group affect, in order to (1) tune the annotation procedure with respect to the social construct and dataset at hand, and (2) train our annotators. For this pilot study, we handpicked three videos from the MeMo corpus, the ones which were perceived to have the maximum affective variances. The three videos were then randomly distributed amongst four of the eight annotators to annotate for dynamic group affect using INTERACT \cite{mangold2018discover}. 

The four annotators used for the pilot study were asked to annotated for group affect using the circumplex model \cite{russell1980circumplex}, independently for the arousal and valence dimensions of affect, similar to \cite{MspConv, recolaDB}. 
To introduce the annotators to the concepts required to annotate, all annotators were individually met with and the definition of dynamic group affect and of the circumplex model (i.e., defining arousal and valence) were discussed. This discussion on the definition of key concepts were done to set a primer for further training of annotators as described in Sec.~\ref{sec:training-annot}. 

\subsubsection{Tuning the Annotation Procedure}\label{sec:tuning-procedure}
We wanted to tune the annotation procedure in terms of two key parameters: (i) the scale to be used to annotate group affect and (ii) the time-window size to be used to aptly capture the affective dynamics. For the initial pilot study, we used the Self Assessment Manikin (SAM) \cite{bradley1994measuring} as the scale to annotate, a common scale used to annotate individual-level arousal and valence \cite{recolaDB}. We used 20\,secs as the size of the time-windows, following \cite{mou2019alone}.


After the four annotators completed the annotations of the videos part of the pilot study, a meeting was setup with the annotators to discuss on the two parameters to be tuned. The discussion during the meeting was specifically on two questions: (i) "Should we increase or decrease or keep the same time-window size?", and, (ii) "Do you accept the scale used? Does it well capture the affective expressions in the dataset?". Based on the consensus reached during the discussion, we made two changes to the initial setup. We reduced the time-window size to 15\,secs and replaced the SAM scale with an ordinal scale ranging from 1 to 9. The decision on the time-window size was taken based on the consensus that the dynamics of group affect in the videos fluctuate rather faster and a smaller time-window size would capture the fluctuations better. Similarly, all the annotators felt that the SAM scale had extreme arousal and valence categories on the scale that were not usually present in the spontaneous interactions. Because of this, the resulting affect annotations had only limited affective variances. To adapt to this, the SAM scale was replaced with an ordinal scale, following existing literature which argues for the ordinal nature of emotions \cite{yannakakis2017ordinal}. The pilot study was repeated iteratively until a consensus was reached amongst annotators to freeze the annotation procedure. In this case, we had to do this two times before the annotation procedure  was frozen with 15\,secs time-windows and the ordinal scale for annotation.

Moreover, during the iterative tuning rounds, a critical problem raised by most of the annotators was that a particular moderator in the MeMo corpus dominated most of the interaction preventing the interaction from being an active discussion amongst all group members. With respect to this, we got rid of all 9 interactions from that particular moderator. After this, the final dataset had in total 35 group interactions. We believe that such rich discussions were possible with the annotators because of their educational background and prior experiences with annotating other social behaviours.

\subsubsection{Training Annotators}\label{sec:training-annot}
The main objective of the training was to help annotators attend to the most important aspect of dynamic group affect, and to treat the construct as less objective as possible.

\noindent\paragraph*{\textit{\textbf{Video Markers and Scale Differences}}} 
A pitfall in using an ordinal scale over the SAM scale is that the annotators do not have a reference affective expression associated with a specific scale item, like the illustrations in SAM. To tackle this problem, we trained the annotators with respect to {video markers} for each of the ordinal scale items. The idea was to associate each of the scale item to a 15\,secs video segment taken from interactions in the MeMo corpus. This video marker based training of annotators was inspired by the nonverbal behavioral anchors for affect expressions developed by Bartel and Saavedra \cite{bartelsaavedra2000_markers}, which were subsequently adopted by other works for the annotation of group affect (e.g., \cite{lehmann2011verbal, lei2015affect}) and mood (e.g., Barsade \cite{barsade2002_ripple}). While Bartel and Saavedra \cite{bartelsaavedra2000_markers} devised a list of nonverbal behaviors for each of the scale item, in this work, to also explain the temporal context, we associate a video segment to each of the scale items.
 

Using the annotations obtained from the tuned annotation procedure, we then selected several candidates for each of the ordinal scale elements and discussed these among all four annotators used in the pilot study using the two definitions provided to the annotators (i.e., definition of dynamic group affect and of the circumplex model). The annotators discussed two key topics: (i) "How do you \textit{aggregate the individual affective} expressions to group affect?" (i.e., on the bottom-up phenomena), and, (ii) "How do you track and \textit{aggregate the dynamic fluctuations} in group affect?" (i.e., on the top-down phenomena). Moreover, owing to the ordinal nature of scale, the {scale differences} between video markers belonging to adjacent scale elements were also discussed. Based on these discussions, one candidate was selected as the emotion marker for a particular scale item. These emotion markers are then used as reference videos to explain the respective scale item. At the end, all other annotators who were not part of the initial pilot studies were also trained using the outcome of the studies and the video markers derived for the ordinal scale.


\noindent\paragraph*{\textit{\textbf{Contribution of the Group Facilitators}}} 
The role and the {contribution of the group facilitator} in the interaction towards the group affective state was also discussed. To capture the unscripted nature of the interaction, we instructed the annotators to treat the moderator as one of the group members and not to perceive them differently from the participants, especially in light of the bottom-up nature of collective group affect. 

\noindent\paragraph*{\textit{\textbf{Focusing on Affective Expressions}}} 
Finally, the annotators were made aware of the concept of \textit{affective expressions}. Following the circumplex model of affective expressions \cite{russell1980circumplex}, they were instructed to only focus on the affect expressed by the group members. For example, when one of the observed group members used sarcasm (i.e.,  a positive valence expression to convey negatively valenced conversational content), the annotators were  instructed to only focus on the affect \textit{expressed} (in this case, positive valence). 

\noindent\paragraph*{\textit{\textbf{Annotation Software Setup and Location}}} 
During training and for the entire annotation procedure that followed, we used INTERACT software \cite{mangold2018discover} which provides a graphical user interface where annotators can scroll through videos to annotate observed behavior directly from an audio or video file. The software allowed us to systematize and synchronize the entire annotation procedure across annotators. INTERACT was set up such that the software requested an annotation for every 15 second interval of each recorded video. Clicking on each segment would play the respective slice of the video. The annotators were allowed to watch each 15 second time-window any number of times. They used their number pad in the keyboard to input a number between 1 and 9 as their ordinal scale annotations, and any wrong input would not be accepted by INTERACT. 

To ensure an appropriate setting without distractions, annotators were asked to come to the laboratory where they were provided with an individual workspace, including a desktop computer with INTERACT installed and a two-screen setup. The 35 interaction videos were provided in a randomised order to each of the annotators, to prevent any potential annotator bias which might occur if all the annotators received the videos in the same order. The videos were given to the annotators in batches of 5 videos to ensure that the annotators followed the order provided. The annotators worked 10-20 hours per week, annotating approximately 2-5 videos per week. 


\begin{table}[t]
\centering
\begin{tabular}{r|c|c}
\toprule
                            & Arousal   & Valence \\ \midrule
Quadratic $\kappa$          & 0.407     & 0.578   \\
Cronbach's $\alpha$         & 0.820     & 0.886   \\
Pearson Correlation (PCC)   & 0.506     & 0.639   \\ \bottomrule
\end{tabular}
\caption{Inter-annotator agreements}
\label{tab:interannot-agree}
\end{table}

\begin{table}[]
\centering
\begin{tabular}{c|cc|cc}
\toprule
                         & \multicolumn{2}{c|}{Arousal} & \multicolumn{2}{c}{Valence} \\ \midrule
\multicolumn{1}{c|}{All} & \multicolumn{2}{c|}{0.407}        & \multicolumn{2}{c}{0.578}        \\ \midrule
Excluded Annotator  & $\kappa$   & $\Delta$     & $\kappa$   & $\Delta$      \\
1                       &     0.402      &    $-$0.005        &     0.573    &  $-$0.005      \\
2                        &     0.398     &     $-$0.009         &     0.572    &   $-$0.006           \\
3                        &     0.416     &     $+$0.009         &     0.581    &   $+$0.003           \\
4                        &    0.402     &    $-$0.005    &     0.575    &   $-$0.002           \\
5                        &    0.412      &   $+$0.005    &     0.567    &   $-$0.011           \\
6                        &    0.401       &  $-$0.006    &     0.598    &   $+$0.020           \\
7                        &    0.399      &   $-$0.008   &     0.569    &   $-$0.009           \\
8                        &    0.385      &   $-$0.022   &     0.582    &   $-$0.004           \\ \bottomrule
\end{tabular}
\caption{Quadratic weighted kappa $\kappa$ scores when a particular annotator is excluded. $\Delta$ is the increase ($+$) or decrease ($-$) in $\kappa$ when the annotator is excluded}
\label{tab:interannot-pariwiseagree}
\end{table}

\subsection{Inter-Annotator Agreement}

The annotation procedure ensured that each observed group discussion received at least six annotations throughout (i.e., six annotations for each 15-second window of each of the 35 group discussions). In comparison, most state-of-the-art individual-level affect recognition datasets have only three to five annotators \cite{busso2008iemocap, recolaDB, MspPod}, with literature on uncertainty modeling suggesting that at least four annotations should be collected for a reliable annotation distribution and its ground-truth consensus \cite{prabhuUncertTAFFC24}. 
To measure the inter-annotator agreement, we used three different metrics: (i) quadratic weighted kappa  ($\kappa$) \cite{mchugh2012interrater}, (ii) Cronbach's alpha ($\alpha$) \cite{CronbachIntGliem2003calculating}, and (iii) Pearson's correlation coefficients (PCC) \cite{freedman2007statistics}. 


\noindent\paragraph*{\textit{\textbf{Quadratic weighted kappa $\kappa$ measure }}}The $\kappa$ measure is a variant of the Cohen's kappa, that is specifically designed to measure agreements in the ordinal scale data \cite{mchugh2012interrater}. The $\kappa$ metric can measure agreements between only two set of data samples, hence all possible annotator pairs were used to calculate $\kappa$ and then averaged to obtain the overall measure. From Table~\ref{tab:interannot-agree}, we observe a $\kappa = $0.407 and $\kappa = $0.610 for arousal and valence, respectively. This indicates a moderate agreement for both arousal and valence dimensions of group affect \cite{mchugh2012interrater}. As a comparison, the MSP-Conversation dataset \cite{MspConv}, on a much simpler annotation task of individual-level affect, has an agreement of $\kappa = $0.50 and $\kappa = $0.54 for arousal and valence, respectively (i.e., higher than ours on arousal and lower on valence). 

\noindent\paragraph*{\textit{\textbf{Cronbach's  $\alpha$ measure }}} The $\alpha$ measure is advantageous as it can measure the agreement between an arbitrary number of annotators. The measure accounts for similarity in both the magnitude and the temporal trends in annotation values. From Table~\ref{tab:interannot-agree}, we observe an $\alpha$ of 0.820 and 0.886 for arousal and valence, respectively, indicating a good and satisfactory level of agreement \cite{CronbachIntGliem2003calculating}. For comparison, the annotations of group affect collected in \cite{mou2019alone} have an $\alpha$ agreement of 0.80 for arousal and 0.89 for valence, which is lower than our arousal annotations and virtually the same as our valence annotations. Note here that \cite{mou2019alone} operationalised group affect in a much simpler social settings, without accounting for interactions between interlocutors.


\noindent\paragraph*{\textit{\textbf{Pearson Correlation (PCC) measure }}}As we annotate a time-dependent dynamic construct we also use the PCC measure as the inter-annotator agreement metric. From Table~\ref{tab:interannot-agree}, we observe a PCC of 0.506 and 0.639 for arousal and valence, respectively. The individual-level affect annotations in the RECOLA dataset \cite{recolaDB}, in comparison, have a PCC agreement of 0.435 for arousal and 0.407 for valence, which is much less than our agreement scores, despite being on a much more simpler social construct.



From Table~\ref{tab:interannot-agree} we note that in general the agreement for valence is higher than that of arousal, which aligns with the literature \cite{MspConv, prabhuUncertTAFFC24}. Furthermore, in Table~\ref{tab:interannot-pariwiseagree}, we present the $\kappa$ scores when a particular annotator was excluded, where $\Delta$ is the increase ($+$) or decrease ($-$) in $\kappa$ when the annotator was excluded. Table~\ref{tab:interannot-pariwiseagree} reveals that in most cases (i.e., for six out of eight annotators), excluding the annotator only decreases the $\kappa$ score ($-$) in terms of both arousal and valence. Only for annotator $5$ a large increase in $\Delta$ is noted when excluded, i.e., $\Delta$ of $+0.020$ is noted for valence. In all other cases, the $\Delta$ is rather minimal, i.e., $\Delta<+0.01$. This points to the reliability of the collected annotations, even when annotations from an annotator are omitted.



Previous research points to the value of retaining extra annotations, even if they have less agreement \cite{MspConv, prabhuUncertTAFFC24}, especially in cases of time-continuous annotations. Owing to this, for further experiments we hold all annotations, and for the final ground-truth calculation we will take in to account the annotator-wise correlations, i.e., weighing less agreeing annotators less (similar to \cite{grimm2005evaluation, recolaDB, MspConv}).

\subsection{Ground-truth}

To derive the final ground-truth of dynamic group affect, we use the Evaluator Weighted Estimator (EWE) \cite{grimm2005evaluation}, a common technique to derive ground-truth for individual-level affect recognition \cite{recolaDB, MspConv}. The EWE, to derive the ground-truth, weights annotations with respect to the annotator-wise correlation coefficients, and is formulated for a time segment $t$ in an interaction sample $s$ as follows:
\begin{equation}
    {y}_{s, t}^{\text{EWE}, (i)} = \dfrac{1}{\sum_{k=1}^{K}r_{k, s}^{(i)}}  \sum_{k=1}^{K} r_{k, s}^{(i)} \hspace{0.25cm} y_{s, t}^{(i)}
\end{equation}
where $r_{k, s}^{(i)}$ denotes the average correlation coefficient of the annotator $k$ with other annotators for a particular interaction sample $s$ and emotion dimension $i$. For unreliable annotators with $r_{k, s}^{(i)} < 0$ a lower bound of zero is defined. In cases where all $k$ annotators have the same correlation coefficients $r_{k, s}^{(i)}$, they result in the same correlation weighting and thereby ${y}_{s, t}^{\text{EWE}, (i)} = y_{s, t}^{(i)}$. For the rest of the experiments and analyses in this work we will use ${y}_{s, t}^{\text{EWE}, (i)}$ as the ground-truth of dynamic group affect. 

\begin{figure}[t!]
\centering
\includegraphics[width=0.8\columnwidth]{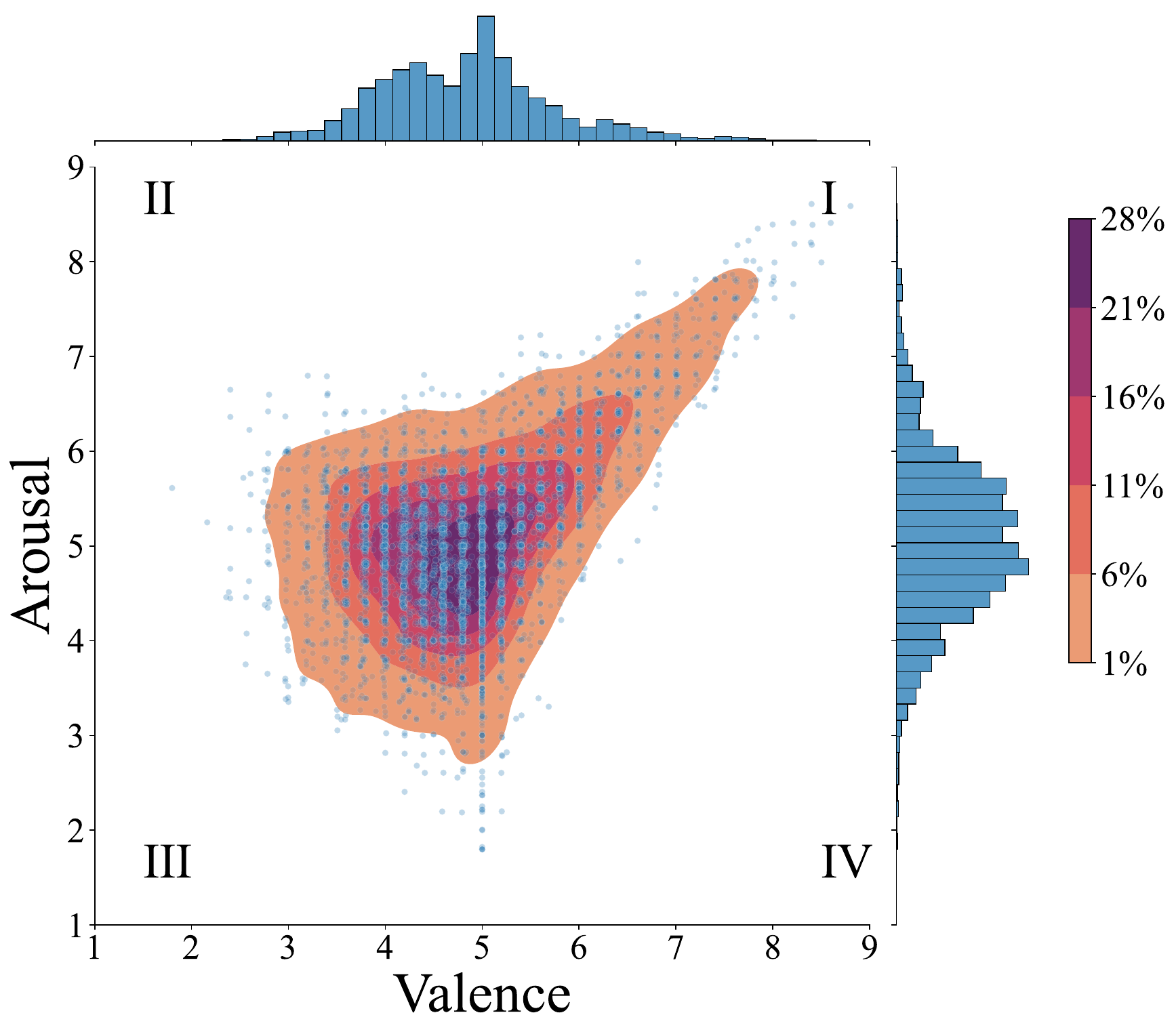}
	\caption{Distribution of final ground-truth for dynamic group affect ${y}_{s, t}^{\text{EWE}, (i)}$, in terms of arousal and valence.}
	\label{fig:gt_dist}%
\end{figure} 

The distribution of the ground-truth, in terms of the arousal and valence dimensions can be seen in Figure~\ref{fig:gt_dist}. From the plot, we note that the independently annotated arousal and valence, depicted by the histograms of the marginal distribution, have considerable variances, i.e., Arousal: $\mu=$ 5.10, $\sigma=$ 0.88, $\min=$ 1.80, $\max=$ 8.61, and, Valence: $\mu=$ 4.84, $\sigma=$ 0.90, $\min=$ 1.80, $\max=$ 8.80.

Samples in Quadrant I denote high-arousal and positive-valence (e.g., emotions such as happy and excitement). Quadrant II denotes high-arousal and negative-valence (e.g., emotions such as anger and frustration). Quadrant III denotes low-arousal and negative-valence (e.g., emotions such as depressed and gloomy). Quadrant IV denotes low-arousal and positive-valence (e.g., emotions such as relaxed). With respect to the joint distribution of arousal and valence, depicted by the heatmap and the scatter plots (in Fig.\ref{fig:gt_dist}), the variances are noted only in some of these quadrants of the circumplex model. For example, good number of samples are noted in Quadrant I. Similarly, a good number of samples can also be noted for low-arousal and neutral-valence (i.e., between Quadrant III and IV; emotions like tired and melancholic), and for neutral-arousal and negative-valence (i.e., between Quadrant II and III; emotions like bored and sad). However, in Quadrant IV and Quadrant II extreme samples are not noted. This could be because the participants in MeMo were reported to be non-preacquainted and complete strangers at the beginning of the longitudinal study that spanned over 3 interactions in two weeks. In such short longitudinal cases, amongst non-preacquainted participants, extreme expressions are very rare \cite{lei2014_groupaffect}, likely due to professional behavioral norms.





\section{Modeling Group Affect Dynamics} \label{sec:modeling}
In this section, we focus on the modeling of dynamic group affect using features that capture the dynamic fine-grained interpersonal relationships behind the emergence of group-level affect. Firstly, in Section~\ref{subsec:featureextract}, we summarise the feature extraction techniques. Secondly, in Section~\ref{subsec:convergedivergence}, we present quantitative analyses on the interpersonal relationship based convergence and divergence phenomena. Finally, in Section~\ref{subsec:predmodel}, we present predictive modeling on the collected group affect annotations


\subsection{Feature Extraction} \label{subsec:featureextract}
For the multimodal modeling of dynamic group affect both audio-based and video-based features are used in this work. The MeMo corpus provides with manually diarized and synchronised audio for each of the interlocutors, collected at a sample rate of 16kHz. Similarly, video recordings of online group discussions are also provided at a frame rate of 60\,fps. The group discussion video frames are cropped to obtain individual-level frames of each of the interlocutors. The summary of the features extracted can be seen in Table~\ref{tab:hand-craft-feats}.


\begin{table}[]
\centering
\begin{tabular}{ll|ll|l}
\toprule
\multicolumn{2}{c|}{\multirow{2}{*}{\begin{tabular}[c]{@{}c@{}}\textbf{Individual}-level\\ features\end{tabular}}} & \multicolumn{2}{c|}{\textbf{Dyad}-level features}                                                                                                                                                                                                                                      & \multicolumn{1}{c}{\multirow{2}{*}{\begin{tabular}[c]{@{}c@{}}\textbf{Group}-level \\ features\end{tabular}}}                                                          \\
\multicolumn{2}{c|}{}                                                                                     & \multicolumn{1}{c|}{Synchrony}                                                                                                                                  & \multicolumn{1}{c|}{Convergence}                                                                        & \multicolumn{1}{c}{}                                                                                                                                          \\ \midrule
\multicolumn{2}{l|}{\textbf{Audio}}                                                                       & \multicolumn{1}{l|}{\multirow{10}{*}{\begin{tabular}[c]{@{}l@{}}(i) correlation \\\vspace{0.1cm} coefficient\\ (ii) lagged \\\vspace{0.1cm} correlation\\ (iii) best lag\end{tabular}}} & \multirow{10}{*}{\begin{tabular}[c]{@{}l@{}}\vspace{0.1cm}(i) global\\ \vspace{0.1cm}(ii) symmetric\\ (iii) asymetric\end{tabular}} & \multirow{10}{*}{\begin{tabular}[c]{@{}l@{}} \vspace{0.1cm}(i) mean\\ \vspace{0.1cm}(ii) deviation\\ \vspace{0.1cm}(iii) median\\ \vspace{0.1cm}(iv) minimum\\ \vspace{0.1cm}(v) maximum\\ (vi) gradient\end{tabular}} \\
1    & Pitch  & \multicolumn{1}{l|}{} &   &           \\
2    & Speech rate  & \multicolumn{1}{l|}{}  &   &    \\
3    & Intensity   & \multicolumn{1}{l|}{}   &   &    \\
4    & MFCC        & \multicolumn{1}{l|}{}   &   &    \\
5    & VGGish     & \multicolumn{1}{l|}{}    &   &    \\  \cmidrule{1-2}
\multicolumn{2}{l|}{\textbf{Video}} & \multicolumn{1}{l|}{}    &     &   \\
1    & Action units  & \multicolumn{1}{l|}{}  &  &     \\
2    & Face pose    & \multicolumn{1}{l|}{}  &  &     \\
3    & ResNet50     & \multicolumn{1}{l|}{}  &  &      \\ 
\bottomrule                                 
\end{tabular}
\caption{Summary of the extracted features.}
\label{tab:hand-craft-feats}
\end{table}

\subsubsection{Audio Features}
As the audio features we extract the first 5 \textit{MFCCs} coefficients, \textit{Voice Intensity}, \textit{Pitch}, \textit{VGGish}, and the \textit{Speech Rate}. This set of individual-level paralinguistic audio cues are selected owing to their demonstrated effectiveness in the automatic recognition of individual-level affect \cite{eyben2015geneva} and also other group-level constructs such as cohesion \cite{Nanninga2017-bi}. The MFCC and pitch features were calculated using the \texttt{librosa}\cite{McFee2015librosaAA} package, for every 10\,ms with a sliding window of 30\,ms. The voice intensity and speech rate features were extracted using \texttt{Praat}\cite{parselmouth}. The voice intensity was calculated at the same rate as the MFCCs while the speech rate was calculated in vowels per second using \cite{de2009praat} at a rate of 1.5\,secs following \cite{Nanninga2017-bi}. The VGGish features are pretrained deep learning based features and was extracted using the pretrained weights from \cite{hershey2017cnn}.


\subsubsection{Video Features}
As the video features we extract \textit{Facial Action Units (AUs)}, \textit{Face Pose}, and \textit{ResNet50}. Action units, Face pose and ResNet50 features have been successfully used in existing literature for several tasks, such as sentiment analysis \cite{tian2001recognizing} and emotion recognition \cite{baltruvsaitis2016openface}. Individual-level AUs (subset available in the \texttt{OpenFace} toolkit) and the face pose (pitch, roll and yaw) were extracted using the \texttt{OpenFace} toolkit \cite{baltruvsaitis2016openface}, for every 0.5\,sec. Similar to the VGGish audio features, the ResNet50 network was used to extract framewise pretrained deep learning based features and was extracted using the pretrained weights from \cite{he2016deep}.

\subsubsection{Group-level Features}
Social interactions are multilevel systems where interpersonal relationships and affective states emerge at multiple levels of the interaction, i.e., at the individual, dyadic and group level \cite{kozlowski2000multilevel}. With respect to this theoretical framework of group-level constructs, in this work, to study dynamic group affect, from the individual-level features we extract dyadic-level and group-level features that are descriptive of the interpersonal relationships shared between a dyad in the interaction and the group as a whole, respectively. 

At the dyad-level, we extract two sets of interpersonal relationship-based features: (1) Synchrony and (2) Convergence. Following \cite{ramanrajprabhu2023}, we use linear correlation coefficient-based features as the synchrony measures: (i) the correlation coefficient $\rho$, i.e., linear correlation without a time-lag, (ii) lagged correlation $\rho_\delta$, the linear correlation with the best time-lag, and (iii) the best lag $\delta$ defined as the time-lag used to obtain the maximum linear correlation between the two individual-level signals. Existing literature reveals that synchronous behaviour is displayed by interlocutors often in a time-lagged manner, with a leader and a follower \cite{delaherche2012interpersonal, ramanrajprabhu2023}. The three measures are formulated as follows:

\begin{equation}
    \begin{split}
        \text{Correlation coeff. $\rho$ : } & X \circledast Y \\
        \text{Lagged correlation $\rho_{\delta}$ : } & \max z(X, Y)  \\
        \text{Best lag $\delta$ : } & \argmax_{l} z(X, Y, l) - ||X|| + 1
    \end{split}
    \label{eq:synchrony}
\end{equation}
\begin{equation}
    z(X, Y, l) = \sum_{k=0}^{||X||-1}X_l \circledast Y_{k-l+N-1} \\    
\end{equation}

where z(., .) is the cross-correlation function, $\circledast$ denotes linear correlation between two signals, $||X||$ denotes the length of signal $X$, $l = 0, 1, ..., ||X|| + ||Y|| - 2$ denoting the time-lags possible, and $N = \max(||X||, ||Y||)$.

As the convergence features, following \cite{Nanninga2017-bi}, we extract (i) global, (ii) symmetric and (iii) asymmetric convergence. Global convergence captures the change in similarity between two individual's social signals, specifically between the initial time-segments and the later time-segments. Similarly, symmetric and asymmetric convergence features capture the decrease or increase in similarity between the two individual's social signals, without and with a time-lag, respectively. The three measures are formulated as below:

\begin{equation}
    \begin{split}
        \text{Global $\Theta_{\text{gbl}}$: } & \sum_{i=0}^{||X||/2} (X_i - Y_i)^2 - \! \sum_{j=||X||/2}^{||X||}(X_j - Y_j)^2 \\
        \text{Symmetric $\Theta_{\text{s}}$: } & (X_l - Y_l)^2 \circledast L \\
        \text{Asymmetric $\Theta_{\text{as}}$: } & p(Y_b/\theta_{X_{a}}) \circledast L
    \end{split}
    \label{eq:convergence}
\end{equation}
where, $L=[0, 1, ..., ||X||]$, $l\in L$, and $\theta$ is the parameter of a Gaussian mixture model (GMM) trained using the expectation-maximization procedure on the data points from $X$ in the initial period of the interaction, i.e., $a\in[0, m], m=2 \cdot ||X||/3$. Similarly, $Y_b$ are data points from $Y$ in the later period of the interaction, i.e., $a\in[m, ||Y||]$.



To extract group-level features from the individual and dyadic features, following \cite{Nanninga2017-bi, ramanrajprabhu2023}, we use six aggregation techniques that are agnostic to the group size. The aggregation measures used are \texttt{average}, \texttt{standard-deviation}, \texttt{median}, \texttt{minimum}, \texttt{maximum}, and \texttt{gradient}.

\subsection{Analysis on Affective Convergence and Divergence} \label{subsec:convergedivergence}

\begin{table*}[t!]
\centering
\begin{tabular}{@{}cc|ccccc|ccccc@{}}

&  & \multicolumn{5}{c|}{\textbf{Arousal}}  & \multicolumn{5}{c}{\textbf{Valence}}  \\ \cmidrule(l){3-12} 

&  & \multicolumn{4}{c|}{Regression Analysis} & Kendal's & \multicolumn{4}{c|}{Regression Analysis} & Kendal's \\

&  & $\alpha$ & $\beta$ & c      & \multicolumn{1}{c|}{$R^2$}    & $\tau$         & $\alpha$ & $\beta$ & c      & \multicolumn{1}{c|}{$R^2$}     & $\tau$         \\ \midrule

\multicolumn{1}{c|}{\multirow{2}{*}{Pitch}}  & $\sigma$    & -2.038  & +22.322  & -12.183 & \multicolumn{1}{c|}{$1.2\%$} & $-0.081^{}$ & -0.650      & -6.654  & +63.554 & \multicolumn{1}{c|}{$1\%$} & $-0.002^{}$  \\
\multicolumn{1}{c|}{}    & $\rho$        & +0.002   &  -0.019   &  +0.023  & \multicolumn{1}{c|}{$6\%$}   &  $+0.210^{*}$    & +0.002   & -0.013  & +0.001 & \multicolumn{1}{c|}{$4\%$}       &  $+0.212^{*}$    \\ \midrule

\multicolumn{1}{c|}{\multirow{3}{*}{VGGish}}                                                & $\sigma$    &  -0.090   &  +0.851  & +0.646 & \multicolumn{1}{c|}{$2.1\%$}      & $-0.105^{}$  &   -0.027    & -0.462   &  +4.168  &  \multicolumn{1}{c|}{$3.7\%$}   & $-0.194^{*}$   \\
\multicolumn{1}{c|}{} & $\rho$      & +0.013   & -0.146  &  + 1.123 & \multicolumn{1}{c|}{$7.2\%$}      &  $+0.027$  & -0.005   &  + 0.018   &  + 0.660 & \multicolumn{1}{c|}{$2\%$}       &   $+0.171^{*}$   \\ 
\multicolumn{1}{c|}{}                                                                       & $\Theta_{\text{as}}$      & +0.030   &  -0.026   &   +0.324     & \multicolumn{1}{c|}{$13\%$}      &  $+0.112^{}$  & -0.001   & +0.018        &   +0.211     & \multicolumn{1}{c|}{$18\%$}       &   $+0.168^{*}$  \\ \midrule

\multicolumn{1}{c|}{\multirow{3}{*}{\begin{tabular}[c]{@{}c@{}}MFCC \\ (1st)\end{tabular}}} & $\sigma$    & -10.925  &    +106.629     & +22.751    & \multicolumn{1}{c|}{$7.7\%$}      &  $-0.170^{*}$     & -3.957   &  26.370  & 242.151   & \multicolumn{1}{c|}{$6.4\%$}       &  $-0.210^{*}$    \\
\multicolumn{1}{c|}{}                                                                       & $\rho_\delta$ & +0.010   &  +0.029       &   +0.001     & \multicolumn{1}{c|}{$16.4\%$}      &  $+0.293^{*}$  & +0.021   & +0.000        &   +0.017     & \multicolumn{1}{c|}{$15\%$}       &   $+0.270^{*}$ \\ 
\multicolumn{1}{c|}{}                                                                       & $\Theta_{\text{s}}$      & +0.001   &  +0.008   &   -0.022     & \multicolumn{1}{c|}{$2\%$}      &  $+0.023^{}$  & +0.003   & -0.034        &   +0.094     & \multicolumn{1}{c|}{$2\%$}       &   $+0.037^{}$  \\ \midrule

\multicolumn{1}{c|}{\multirow{2}{*}{AU06}}      & $\sigma$    & -0.002   & -0.010 &   +0.058  & \multicolumn{1}{c|}{$3.3\%$}     &      $-0.210^{*}$   & -0.001   &   -0.004 &  +0.077   & \multicolumn{1}{c|}{$6.6\%$}       &  $-0.259^{*}$     \\
\multicolumn{1}{c|}{} & $\rho$ & +0.018   &  -0.125   &  +0.209 & \multicolumn{1}{c|}{$15.5\%$}   &  $+0.264^{*}$       & +0.017   &  -0.102  &  +0.150  & \multicolumn{1}{c|}{$18.5\%$}       &   $+0.275^{*}$     \\ \midrule

\multicolumn{1}{c|}{\multirow{2}{*}{AU07}}                                                  & $\sigma$    & +0.002   & + 0.001 &   +0.067  & \multicolumn{1}{c|}{$7.0\%$}      &      $-0.213^{*}$   & -0.002   &   0.025 &  -0.004   & \multicolumn{1}{c|}{$14.8\%$}       &  $-0.343^{*}$     \\
\multicolumn{1}{c|}{}   & $\rho$ & +0.077   &  -0.059  &  +0.107  & \multicolumn{1}{c|}{$2.0\%$}      &    $+0.138$        & +0.067   &  -0.047  &  +0.076 & \multicolumn{1}{c|}{$2.0\%$} & $+0.141$     \\ \midrule

\multicolumn{1}{c|}{\multirow{2}{*}{AU12}}                                                  & $\sigma$    & -0.004   &   -0.007  &  +0.277  & \multicolumn{1}{c|}{$1.5\%$}      &  $-0.104$   & -0.003   &  +0.02    &  +0.211   & \multicolumn{1}{c|}{$2.2\%$}       &  $+0.130^{}$   \\
\multicolumn{1}{c|}{} & $\rho$ & +0.019   &  -0.135  &  +0.240  & \multicolumn{1}{c|}{$16.1\%$}      &   $+0.230^{*}$  & +0.015   &  -0.085  &  0.115  & \multicolumn{1}{c|}{$18.4\%$}       &   $+0.272^{*}$  \\ \midrule

\multicolumn{1}{c|}{\multirow{2}{*}{AU25}}                            & $\sigma$    & -0.012   &  +0.009 &  +0.135 & \multicolumn{1}{c|}{$3.1\%$}      &  $+0.027^{}$    & -0.004   & +0.069    &  -0.118   & \multicolumn{1}{c|}{$3.5\%$}       &  $+0.233^{*}$          \\
\multicolumn{1}{c|}{}  & $\rho$ & +0.013   &  -0.109  & +0.224 & \multicolumn{1}{c|}{$8\%$} & $+0.125$     &  +0.012  & -0.090   & +0.172 &  \multicolumn{1}{c|}{$8\%$}       &    $+0.132$     \\ \midrule

\multicolumn{1}{c|}{Head Roll}  & $\rho_\delta$ & +0.178   &  -1.799  & +14.633  & \multicolumn{1}{c|}{$2\%$}      &  $+0.015$    & +0.113   &  -1.243   & 13.511   & \multicolumn{1}{c|}{$2\%$}       &   $+0.035$     \\ \bottomrule
 
\end{tabular}
\caption{Quantitative analysis of the convergence-divergence processes.}
\label{tab:convdiv-quant}
\end{table*}


\begin{figure*}[t!]
     \captionsetup[subfigure]{justification=centering}
     \centering
     \begin{subfigure}[b]{0.238\textwidth}
            \includegraphics[width=\textwidth]{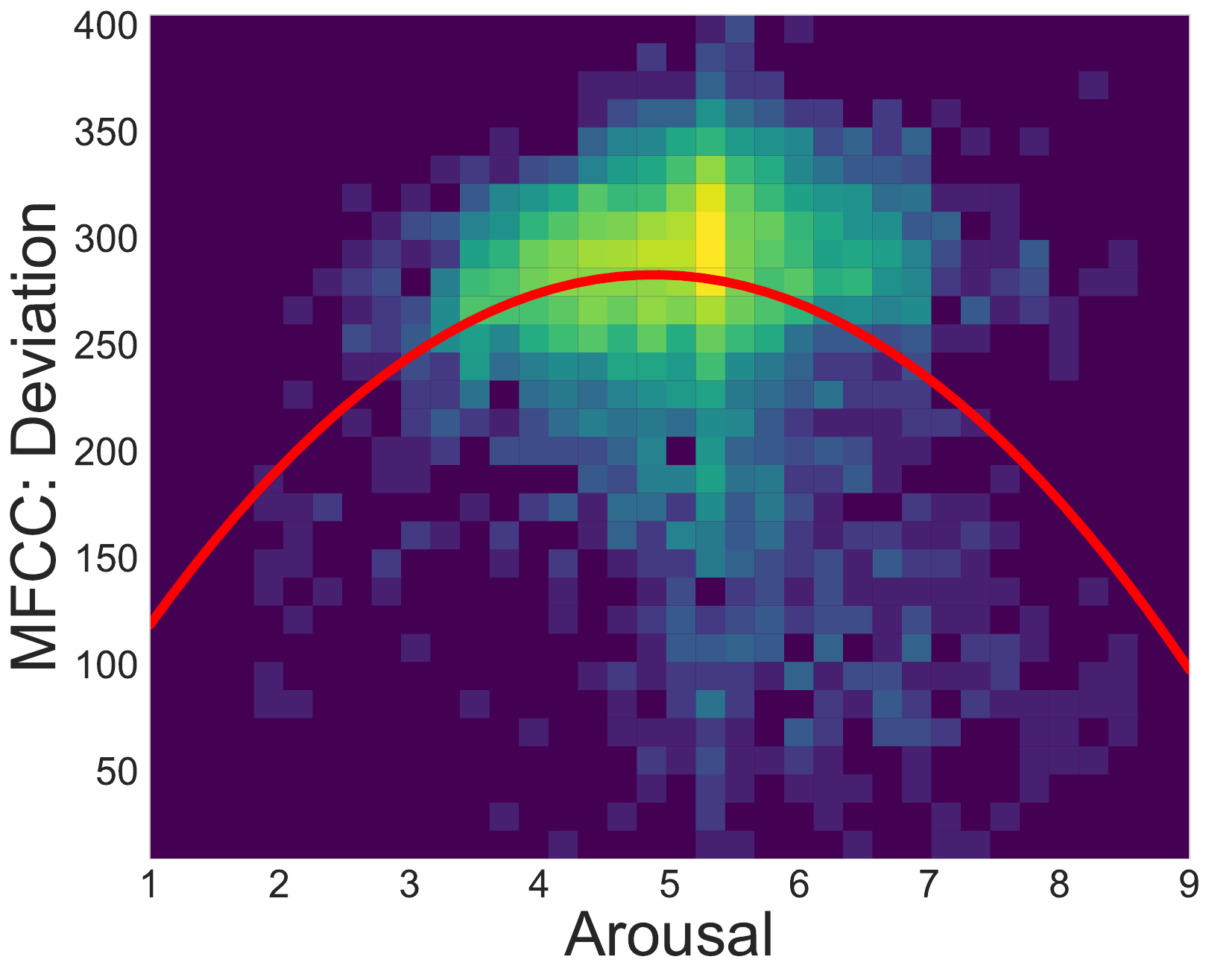}%
         \caption{Standard deviation}
      \label{fig:mfcc-std-arousal}%
     \end{subfigure}
     \hfill
     \begin{subfigure}[b]{0.238\textwidth}
            \includegraphics[width=\textwidth]{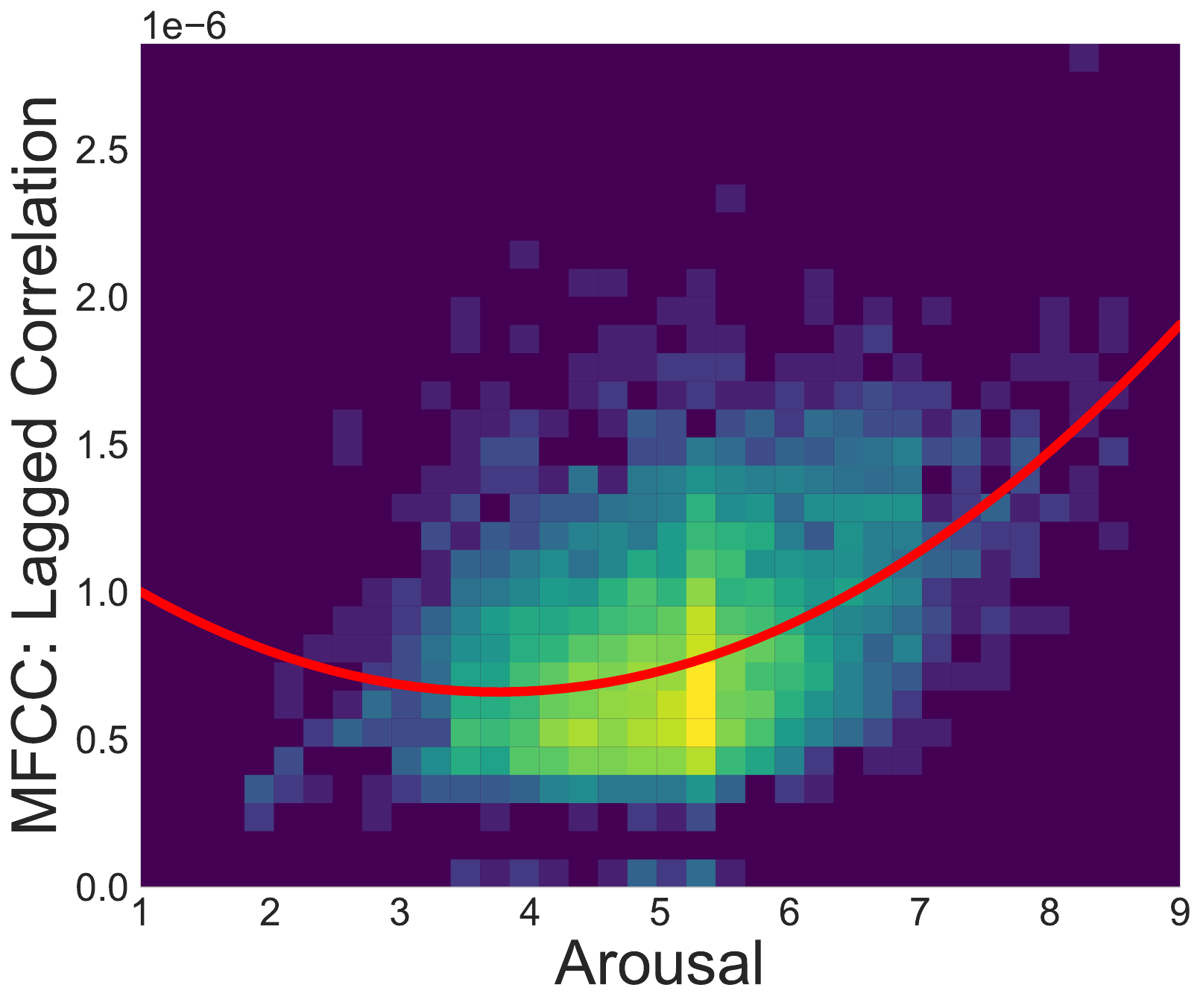}
         \caption{Time-lagged correlation}
      \label{fig:mfcc-lagcorrel-arousal}%
     \end{subfigure}
     \hfill
     \begin{subfigure}[b]{0.238\textwidth}
            \includegraphics[width=\textwidth]{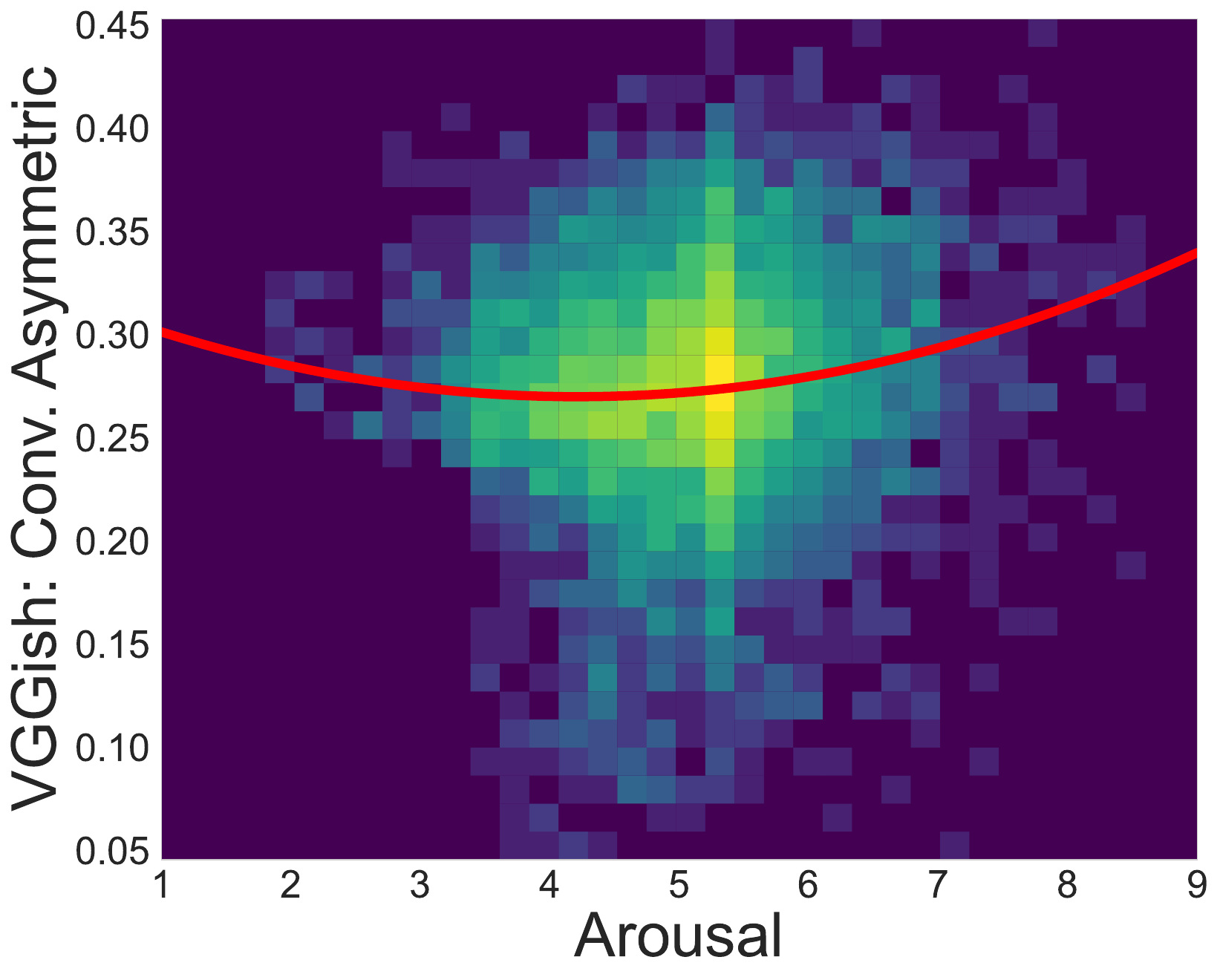}
         \caption{Asymmetric Convergence}
      \label{fig:vggish-asymmconv-arousal}%
     \end{subfigure}
     \hfill
     \begin{subfigure}[b]{0.266\textwidth}
            \includegraphics[width=\textwidth]{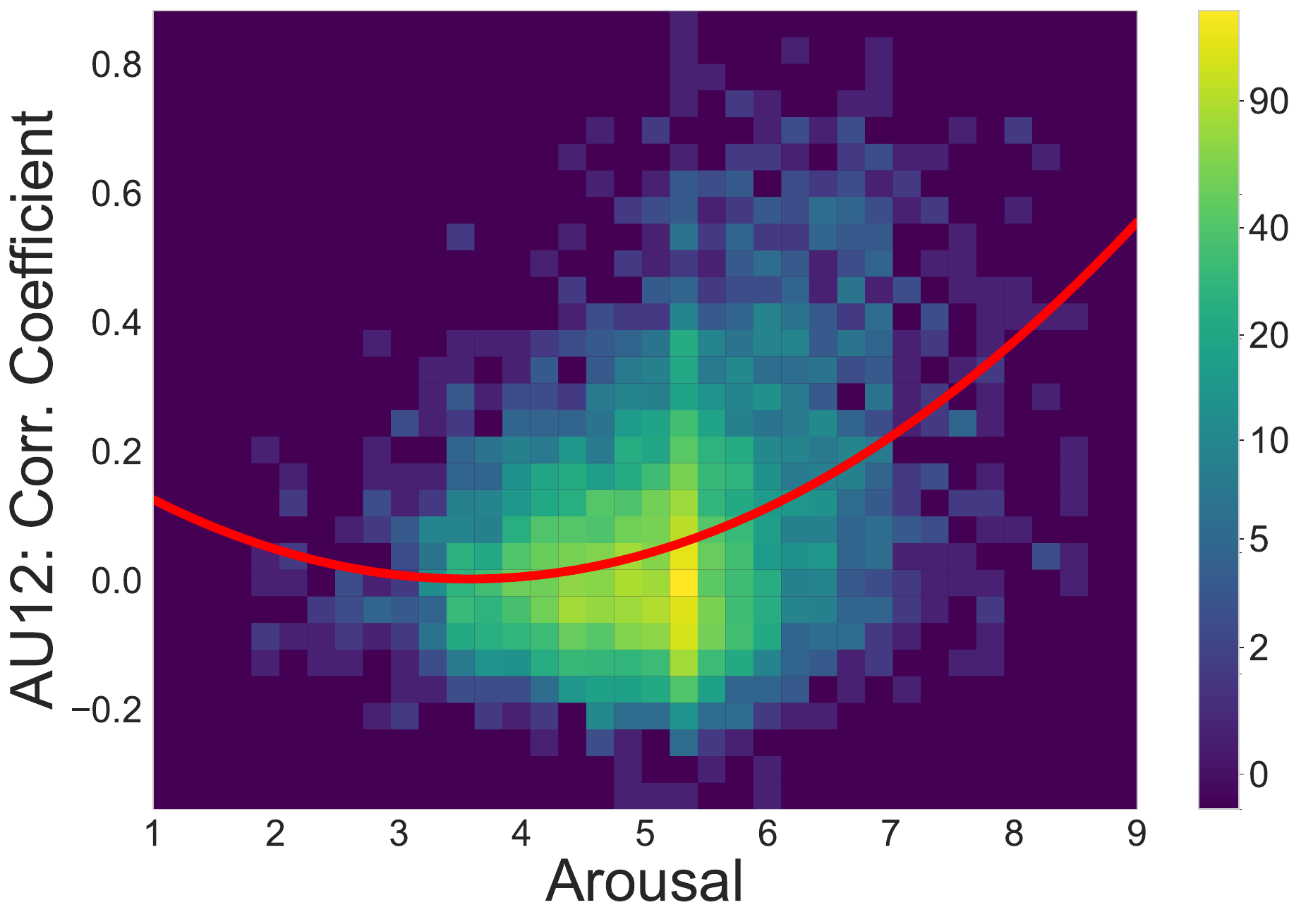}
         \caption{Correlation coefficient}
         \label{fig:au12-corrcoeff-arousal}
     \end{subfigure}
        \caption{Relationship between convergence-divergence measures and group affect in terms of \textit{Arousal}}
        \label{fig:condiv-plot-arousal}
\end{figure*}

\begin{figure*}[t!]
     \captionsetup[subfigure]{justification=centering}
     \centering
     \begin{subfigure}[b]{0.238\textwidth}
            \includegraphics[width=\textwidth]{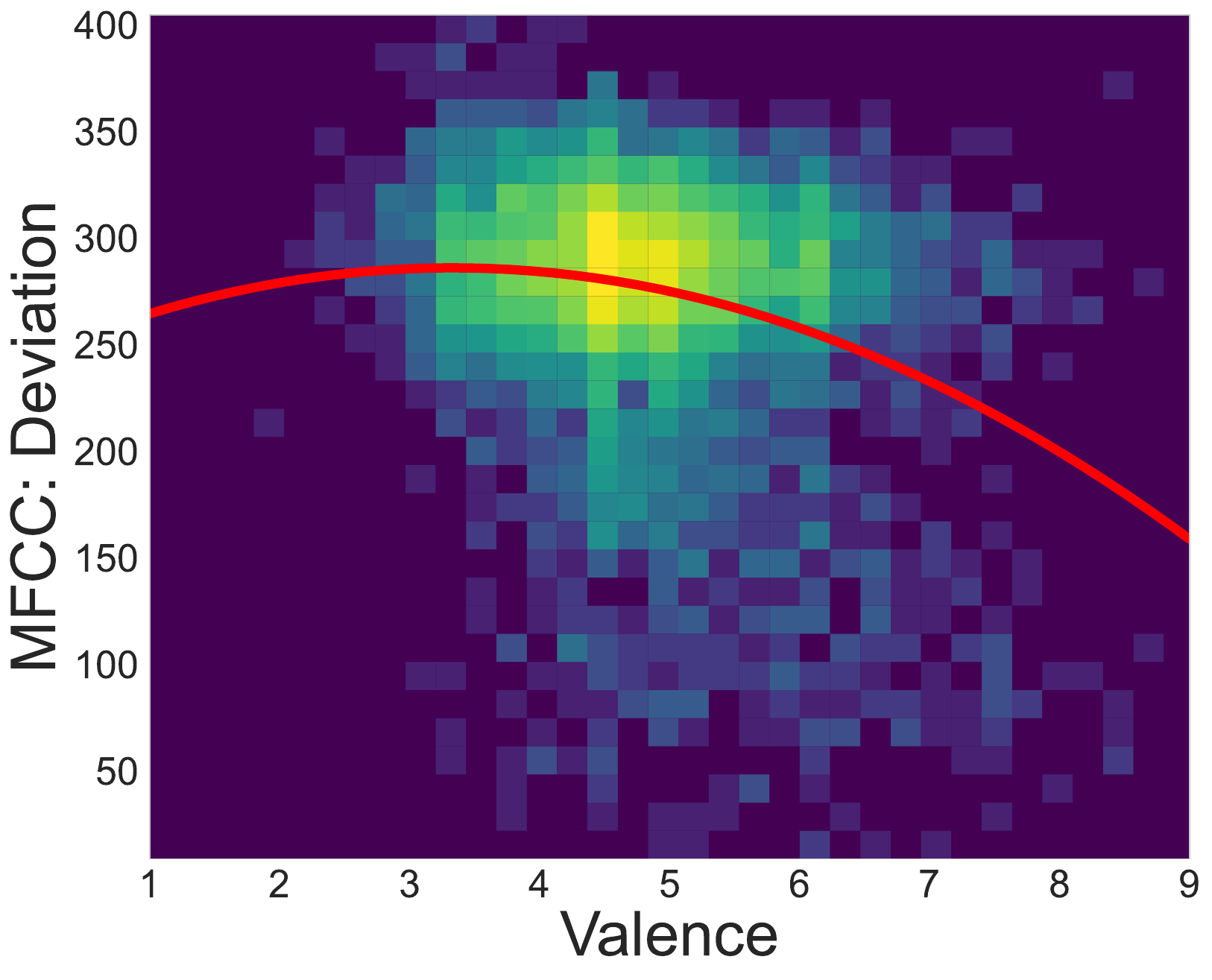}%
         \caption{Standard deviation}
      \label{fig:mfcc-std-valence}%
     \end{subfigure}
     \hfill
     \begin{subfigure}[b]{0.238\textwidth}
            \includegraphics[width=\textwidth]{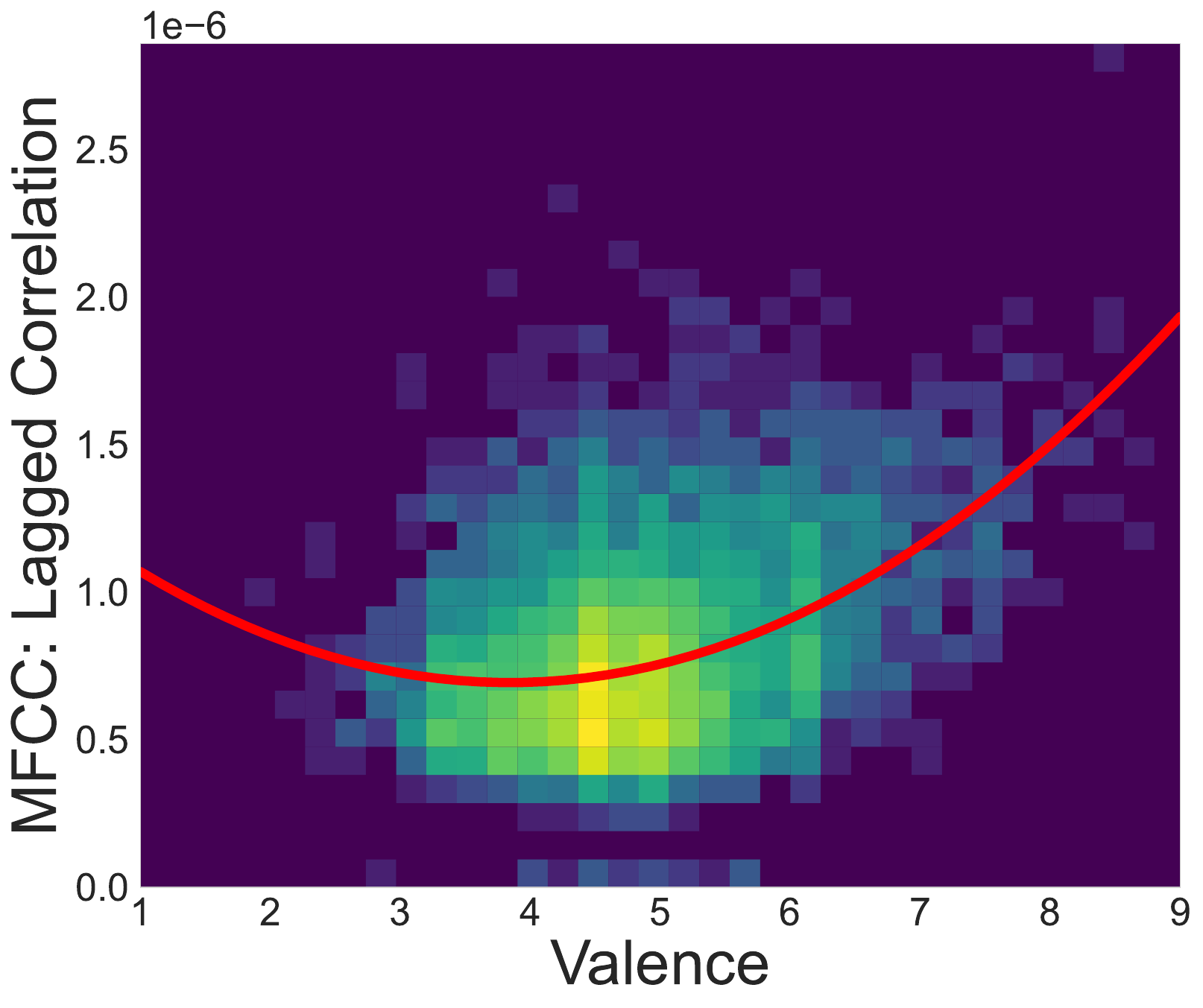}%
         \caption{Time-lagged correlation}
         \label{fig:mfcc-lagcorrel-valence}
     \end{subfigure}
     \hfill
     \begin{subfigure}[b]{0.238\textwidth}
            \includegraphics[width=\textwidth]{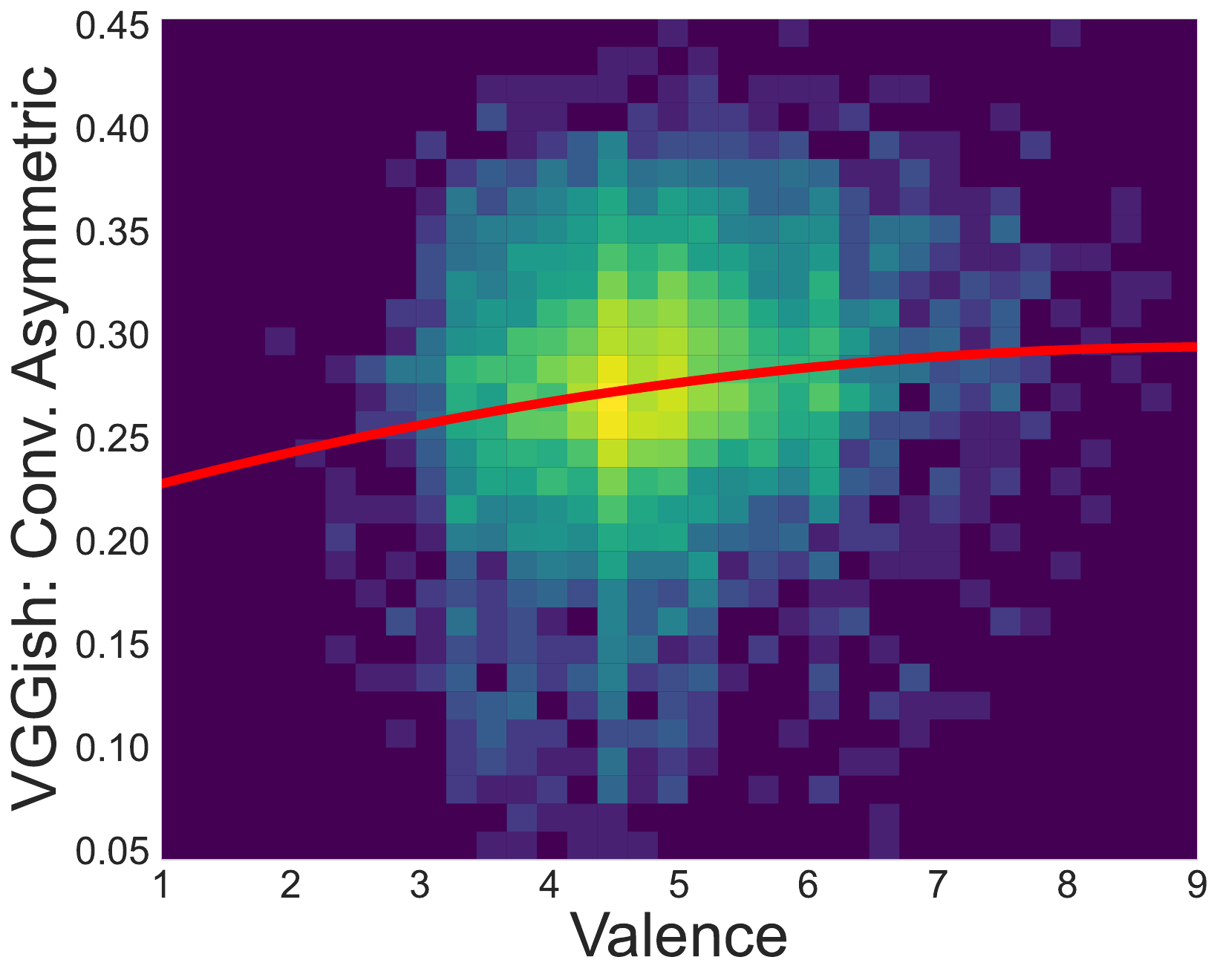}%
         \caption{Asymmetric Convergence}
      \label{fig:vggish-asymmconv-valence}%
     \end{subfigure}
     \hfill
     \begin{subfigure}[b]{0.266\textwidth}
            \includegraphics[width=\textwidth]{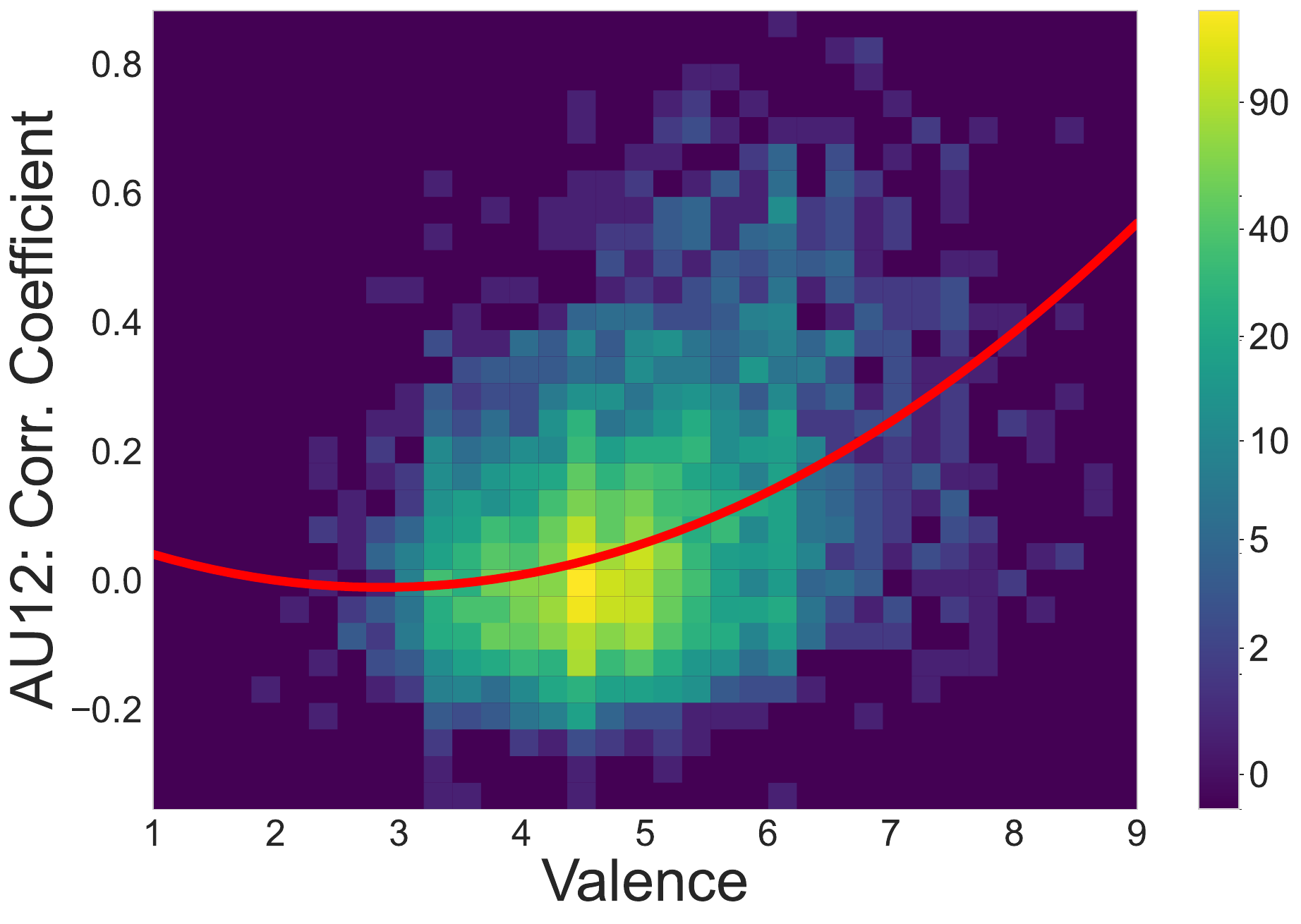}%
         \caption{Correlation coefficient}
         \label{fig:au12-corrcoeff-valence}
     \end{subfigure}
        \caption{Relationship between convergence-divergence measures and group affect in terms of \textit{Valence}}
        \label{fig:condiv-plot-valence}
\end{figure*}

With the majority of empirical research focused on static group affect \cite{dhall2015more, dhall2017GAchallenge, wang2023implementing}, Kelly \& Barsade \cite{kelly2001mood} emphasized on the dynamic nature of affect, i.e., how, over time, the nature of collective affect can change. This ebb-and-flow of affect in groups over time is primarily characterised by the affective convergence and divergence underlying the bottom-up and top-down processes of group affect \cite{hareli2008emotion, barsade2015group}. Foundational theories on affective dynamics (e.g., \cite{hatfield1994emotional}) further describe how several individual interaction- and behaviour-level mechanisms, including facial mimicry, emotional similarity and dissimilarity, and empathy, contribute to affective \textit{convergence} and \textit{divergence} in groups. 

Building on these theorisations, in this work, we present a quantitative analysis on the relationship between interaction- and behaviour-level cues, that quantify the level of affective convergence and divergence within interlocutors, and the collected annotations of dynamic group affect.



 

\subsubsection{Experimental Setup}
For this experiment, as the \textit{independent variable}, we use the group-level features that explain the within-group convergence and divergence, i.e., the similarity and dissimilarity in social signals amongst interlocutors (based on foundational theories presented by Hatfield et al. \cite{hatfield1994emotional}). For example, from Table~\ref{tab:hand-craft-feats}, all the \texttt{mean} aggregation of the {synchrony} and {convergence} based features (i.e., Eq.~\eqref{eq:synchrony} and Eq.~\eqref{eq:convergence}) and the \texttt{standard-deviation} ($\sigma$) of all the \textit{individual}-level features are used. Intuitively, \textit{larger} $\sigma$ values indicate a group that is \textit{diverging}, while \textit{smaller} values indicate that it is \textit{converging}. Contrarily, \textit{larger} values of the \textit{synchrony} and \textit{convergence} based features imply convergence and \textit{smaller} values that of divergence. As the \textit{dependent variable}, the ground-truth annotations of dynamic group affect is used. 

For qualitative analysis of the relationship we plot a 2-dimensional histogram between the two variables, with the dependent variable of group affect in the x-axis and the respective independent variables in the y-axis. Furthermore, we also use a least-squares based polynomial regression with \textit{two} degrees of freedom, where the relationship is modeled as a $2^{nd}$ degree polynomial in the independent variables. The regression model is formulated as:
\begin{equation}
    y_t = \alpha x_t^2 + \beta x_t + c\,.  
\end{equation}
The regression model $y_t$ is overlayed on the histogram, and can be seen in Figures~\ref{fig:condiv-plot-arousal} and \ref{fig:condiv-plot-valence}.

For the quantitative analysis the regression coefficients of the polynomial model ($\alpha$, $\beta$, and $c$) are analysed for all the independent measures used. Along with the regression coefficients, the polynomial model's R-Squared (R$^2$) is also analysed. Additionally, the Kendal's rank correlation coefficient $\tau$, along with its statistical significance ascertained with a two-tailed p-value $\leq 10\%$ (denoted by $*$), is  used to reveal the direction (positive or negative) of linear relationship in the ordinal scale of group affect. 

\subsubsection{Quantitative Analyses and Discussion}
From the results presented in Table~\ref{tab:convdiv-quant}, and, the plots in Figures~\ref{fig:condiv-plot-arousal} and \ref{fig:condiv-plot-valence}, we observe the following.

\textbf{Trends across the affect scale:} We note that the interacting groups tend to \textit{diverge} in terms of their social signals along neutral levels of arousal and valence (i.e., mid-scale values of 4 to 6) and \textit{converge} along extreme levels of arousal and valence (i.e., strong positive affect values of 8-9, or, strong negative values of 1-2). This trend is inferred using the \textit{negative} $\alpha$ values for deviation based group-level features (i.e., $\sigma$ features), and \textit{positive} $\alpha$ values for synchrony and convergence based features (i.e., $\rho$, $\rho_\delta$, $\Theta_{\text{s}}$ features). Note that negative $\alpha$ values denote concave curves (e.g., seen in Figures~\ref{fig:mfcc-std-arousal} and \ref{fig:mfcc-std-valence}), and positive $\alpha$ values denote convex curves (e.g., seen in Figures~\ref{fig:mfcc-lagcorrel-arousal} and \ref{fig:mfcc-lagcorrel-valence}).
\\\textbf{Positive Vs Negative affect:} We note that the degree of convergence in higher for positive affect than that for negative affect, both in terms of arousal and valence. This is inferred from the \textit{negative} Kendal's $\tau$ for $\sigma$ features, and the \textit{positive} $\tau$ for synchrony and convergence based features. Further inference also reveals that this trend is stronger for valence than for arousal. For example, for MFCC-based deviation features $\sigma$, Kendal's $\tau$ is stronger for valence ($\tau=-$0.210; depicted in Fig.~\ref{fig:mfcc-std-valence}) than for arousal ($\tau=-$0.170; depicted in Fig.~\ref{fig:mfcc-std-arousal}). This is consistent across all other features. In such cases, the strength of the $\alpha$ coefficients of the regression curves reduces along with a deteriorating $R^2$, while the strength of Kendal's $\tau$ increases. That is, the relationship between group affect and convergence/divergence features is more linear in terms of valence, and it is more binomial (concave or convex) in terms of arousal.
\\\textbf{Few features with low $R^2$ and lacking significance:} While the trends discussed above are consistent across features and affect dimensions, we note that for \textit{few features} they are backed by rather low $R^2$ and lacking statistical significance. A probable explanation to this observation is that group affect is a complex social construct that can not be explained well by a single feature. However, in this particular analysis, we specifically wanted to study the convergence-divergence processes with respect to each of these features separately for better explainability. In the subsequent Section~\ref{subsec:predmodel}, we perform predictive modeling of group affect by \textit{jointly} modeling the different sets of features extracted.  
 
\subsection{Predictive Modeling} \label{subsec:predmodel}
In this section, we combine the extracted features (see Table~\ref{tab:hand-craft-feats}) to perform predictive modeling of dynamic group affect. These results will serve as the baselines for future research on modeling the novel annotations collected for dynamic group affect.

\subsubsection{Experimental Setup}
\textbf{Feature Sets:} For the predictive modeling, we used three sets of group-level features: (i) the \textit{Basic} feature set, where the individual-level features are directly aggregated to group-level using the group-level aggregators (e.g., \text{mean}, \text{standard-deviation}, \text{gradient}). Note that this feature set does not capture any interpersonal and dyadic relationships in the interaction. Moreover, they also do not capture the social signal dynamics aptly as they are extracted using both the average temporal aggregation and the group-level aggregations. Secondly, (ii) the \textit{Synchrony} feature set, where dyadic synchrony and convergence based features are extracted before the group-level aggregations. This feature set further captures both the dynamics of social signal and the interpersonal relationships amongst interlocutors. Finally, (iii) \textit{Combined}, where both the Basic and Synchrony feature sets are fused and modeled together. These three feature sets are further categorised as audio, video, and, audio-visual feature sets.\\
\textbf{Data Partition:} The dataset is partitioned following the strategy proposed in \cite{MspPod}. The partitions were made such that there is no speaker overlap between the training and testing datasets. This also includes non-overlapping moderators in the MeMo corpus. The validation dataset however may have an overlap of moderators in some samples, but not overlapping participants. Overall, the training-testing split is made with a 80-20$\%$ split, and 10$\%$ of the training datasets is split for the validation dataset. \\
\textbf{Predictive Model:} To model these features, we use a simple Multi-Layer Perception (MLP). The MLP is made up of three linear layers with ReLU and Batch Norm after each of the layers. The MLP's architecture was tuned with respect to the loss obtained on the validation dataset.\\ 
\textbf{Loss Function:} The concordance correlation coefficient (CCC) \cite{lawrence1989concordance} is used as the loss function and as the metric to validate the performances. The CCC has been widely used in literature for the task of individual-level affect recognition \cite{Schuller2018-xi}. The CCC measures the agreement between two variables and ranges from $-$1 to $+$1, with perfect agreement at $+$1. In contrast to Pearson's correlation, the CCC takes both the linear correlation and the bias in to account, which makes it preferable over Pearson's correlation as the loss function and as the evaluation metric. \\
\textbf{Training Strategy:} The models are trained using the ADAM optimizer with a learning rate of $10^{-4}$. The strategy includes an early-stopping on the validation loss improvements with a patience of 10 epochs. The best model during training is selected as the one with the best validation loss. \\


\subsubsection{Results and Discussion}
                              
                              

\begin{table}[t!]
\begin{tabular}{@{}c|l|c|ccc@{}}
\toprule
                              & \multicolumn{1}{c|}{\begin{tabular}[c]{@{}c@{}}Feature \\ Set\end{tabular}} & \multicolumn{1}{c|}{\begin{tabular}[c]{@{}c@{}}\# Model\\ Parameters\end{tabular}} & \multicolumn{1}{c}{\begin{tabular}[c]{@{}c@{}}Arousal\\ CCC $\uparrow$\end{tabular}} & \multicolumn{1}{c}{\begin{tabular}[c]{@{}c@{}}Valence\\ CCC $\uparrow$\end{tabular}} & \multicolumn{1}{c}{\begin{tabular}[c]{@{}c@{}}Average\\ CCC $\uparrow$\end{tabular}} \\ \midrule
\multirow{3}{*}{Audio}        & Basic  & $\approx$15k & 0.242    & 0.245    & 0.244     \\
                              & Synchrony    & $\approx$46  & 0.215    & 0.203  & 0.209      \\
                              & Combined   & $\approx$52k    & 0.261    & 0.222       & 0.241    \\ \midrule
                              
\multirow{3}{*}{Video}        & Basic     & $\approx$27k    & 0.315    & 0.405   & 0.360   \\
                              & Synchrony   & $\approx$115k   & 0.342     & 0.428  & 0.385    \\
                              & Combined  & $\approx$132k   & 0.355     & 0.448    & 0.401      \\ \midrule
                              
\multirow{3}{*}{\begin{tabular}[c]{@{}c@{}}Audio-\\ Visual\end{tabular}} & Basic  & $\approx$33k   & 0.293  & 0.332   & 0.313    \\
                              & Synchrony   & $\approx$152k  & 0.403    & 0.428   & 0.401    \\
                              & Combined    & $\approx$175k    & 0.416   & 0.431  & 0.445   \\ \bottomrule
\end{tabular}
\caption{Results of the audio-visual predictive modeling. Number of model parameters presented in thousands (k).}
\label{tab:pred-modeling}
\end{table}

The results of the predictive modeling is presented in Table~\ref{tab:pred-modeling}. The following takeaways can be made from the results on the prediction of dynamic group affect. \\
\textbf{Multimodal nature of group affect: } The results reveal that the emergence of group affect can be best captured in a multimodal manner, with the audio-visual feature set obtaining the best performance of $0.416$ and $0.431$ in terms of arousal and valence, respectively. Moreover, between the audio and video modalities, we can note that the video-based feature sets are better predictors of group affect than audio-based feature sets, with better performances across the Basic, Synchrony, and, Combined feature sets. We also note that the audio modality better explains the arousal dimension than the valence dimension, while the video modality better predicts the valence dimension. However, their performance differences are not as large as noted in individual-level affect recognition literature \cite{prabhuUncertTAFFC24}. \\
\textbf{Relevance of capturing interpersonal relationships: } In this work, dynamic group affect was mainly studied using the synchrony and convergence based features. Such features, by the manner of their definition and their extraction capture the interpersonal relationships within groups and the dynamics of social signals that lead to the emergence of dynamic group affect. From the results we note that the inclusion of interpersonal relationship based features improves the predictive performances, by better explaining the emergence of dynamic group affect. For example, the inclusion of such features in the video and audio-visual feature sets improves the average CCC from $0.360$ to $0.401$, and from $0.401$ to $0.445$, respectively. However, in case of the audio modality, the synchrony features perform worse than the basic feature set. We attribute this specific behaviour in the audio modality towards the presence of several segments where most of the interlocutors remain silent, i.e., absence of audio modality hinders the extraction of synchrony features. This is not the case with the video modality as the modality is active for most of the time, e.g., active/attentive listeners \cite{silentparticipant}. With respect to these results, we recommend an audio-visual strategy for extracting synchrony based features for modeling group-level social constructs.


\section{Conclusion}  \label{subsec:conclusion}
The majority of the literature on group affect is agnostic to the inherent dynamic nature of the construct \cite{sharma2021audio, dhall2017GAchallenge, huang2018multimodal, wang2023implementing}, and existing methodologies and modeling techniques have been limited by the lack of dynamic annotations of group affect (e.g., \cite{dhall2015more, wang2023implementing}). Addressing this, in this work, as the first in literature, we collected annotations for group affect by also accounting for its underlying fine-grained temporal dynamics. The annotation procedure deployed used a window size of 15\,secs to capture the dynamics in group affect, and quantified affect in an ordinal scale \cite{yannakakis2017ordinal} with respect to Russel's circumplex model (in terms of arousal and valence) \cite{russell1980circumplex}. The inter-annotator agreement analysis on the collected annotations indicated a moderate level of agreement as per the interpretation of Cohen's $\kappa$. We anticipate and hope that the dynamic group affect annotations collected as part of this work will trigger future research on the emergence of group affect as a collective level construct. 



Following theorizing on group affect from the organizational psychology literature \cite{hareli2008emotion, barsade2015group}, we studied the ebb and flow of dynamic group affect by investigating multimodal affective convergence and divergence among group members across the observed group discussions in the MeMo corpus \cite{memo}. Our quantitative analyses on the relationship between the dynamic interpersonal relationship based features (e.g., interpersonal synchrony and convergence; see Table~\ref{tab:hand-craft-feats}) revealed that the \textit{divergence} of group members' social signals results in neutral levels of group affect, whereas their \textit{convergence} results in extreme levels of group affect (i.e., either strong positive or strong negative affect). We also showed that group members converged more when positive group affect emerges than when negative group affect emerges. Subsequently, our predictive modeling of dynamic group affect revealed that it is important to use interpersonal relationship-based features (i.e., synchrony and convergence) to aptly model group affect. Furthermore, the multimodal nature of group affect was also revealed, with the fusion of audio and video modality yielding more accurate predictions of both arousal and valence.

\subsection{Limitations and Future Avenues}

First, the social interactions present in MeMo \cite{memo} occur among groups of non-preacquainted participants, with a short longitudinal study spanning 3 interactions over the course of 2 weeks. While this zero-history group setup is very suitable to study emergence processes such as group affect \cite{kozlowski2015advancing}, we cannot draw conclusions regarding collective affect and its convergence or divergence mechanisms in \textit{groups that share a history}. Despite our observations of rather vivid discussions among participants in MeMo \cite{memo}, the interactions are still not "real" teams that collaborate on a day-to-day basis, which may explain the relatively small nuances of affective variance in our annotations (see Fig.~\ref{fig:gt_dist}). Hence, it would be of interest to collect group affect annotations on longitudinal data of "real" teams that collaborate on a day-to-day basis as a future research endeavor.



Second, our modeling of group affect involved extracting interpersonal synchrony features independently from both the audio and video signals. In this case, the cross-modal relationship between social signals is not captured directly. For example, an event when a group member raises their eyebrow (captured via video features) as an emotional reaction to a fellow group member's raised voice pitch (captured via audio features) is not captured. In future work, it would be apt to include such cross-modal synchrony features into the convergence and divergence analysis.


%

Third, the results of our predictive modeling of dynamic group affect are preliminary baseline results obtained using simple neural network architectures and hand-crafted features. Future work should strive to improve these baselines with more complex neural networks (e.g., graph networks \cite{wang2023implementing}) and end-to-end feature extractors (e.g., \cite{prabhuUncertTAFFC24}). This would represent a significant step towards aptly modeling dynamic group affect and to automate the task of generating group affect ground-truth to study group processes \cite{nalecohesion2024}.

Finally, we argue for more interdisciplinary efforts in the future to advance both the social signal processing and the group process literature more broadly. For example, designing novel neural network architectures for group affect modeling by drawing domain knowledge from the organizational psychology literature. The development of such techniques could in turn help group scholars to better understand the processes underlying the emergence of affect in groups. Moreover, interdisciplinary future work can investigate how convergence and divergence processes, and the underlying social signal mimicry, may translate across various group phenomena (e.g., \cite{nalecohesion2024}) or show different multimodal behavioral imprints that are specific for different phenomena in dynamic group interactions.


\bibliographystyle{IEEEtran}
\bibliography{references}
\vspace{-1em}
\begin{IEEEbiography}[{\includegraphics[height=1.1in,width=1in,clip,keepaspectratio]{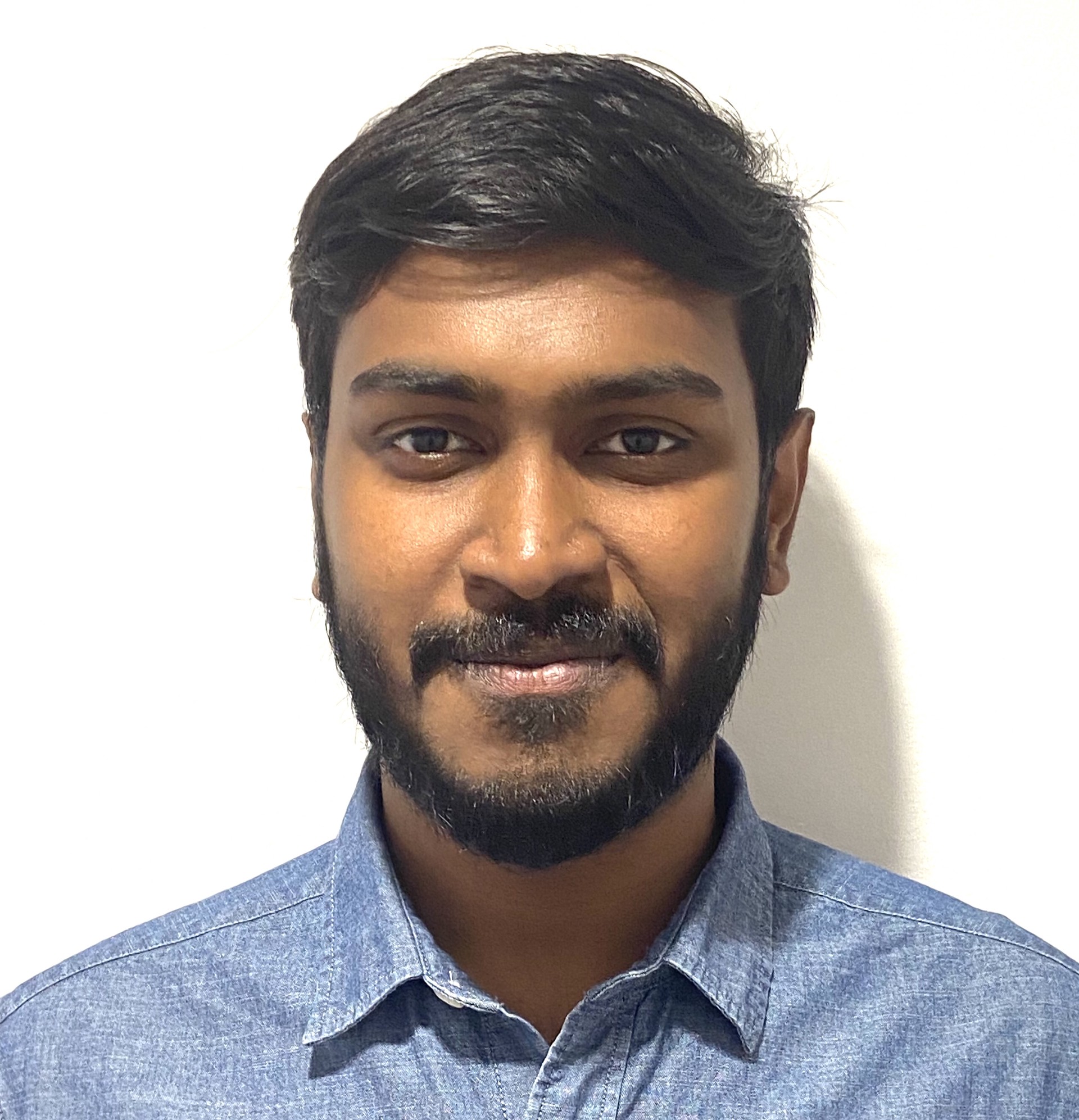}}]{Navin Raj~Prabhu}
received a B.Tech degree in Computer Science from SRM University, India, in 2015, and the MS degree in Computer Science from Delft University of Technology, The Netherlands, in 2020. Currently, he is a PhD student at the Signal Processing Lab and the Organisation Psychology Lab, University of Hamburg, Germany. His research interests include affective computing, social signal processing, deep learning, uncertainty modelling, emotional speech synthesis, and group affect. 
\end{IEEEbiography}

\begin{IEEEbiography}
[{\includegraphics[height=1.2in,clip,keepaspectratio]{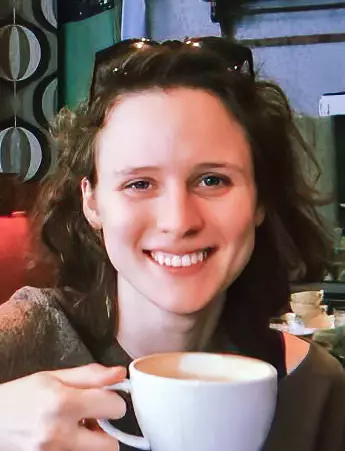}}]{Maria Tsfasman} (Masha) holds a BSc degree in Fundamental and Computational linguistics from HSE university, Moscow and MSc degree in Artificial Intelligence from the Radboud University, Nijmegen with distinction. During her studies, she has done research visits to ISIR,  Sorbonne University, Paris (3 months) and IRCN, the University of Tokyo (8 months). She is currently pursuing a PhD in Delft University of Technology. Her research interests lie between cognitive and computer science: training machines to understand humans better and using the insights from these machines to expand global understanding of human social processing and cognition.
\end{IEEEbiography}

\begin{IEEEbiography}[{\includegraphics[height=1in,clip,keepaspectratio]{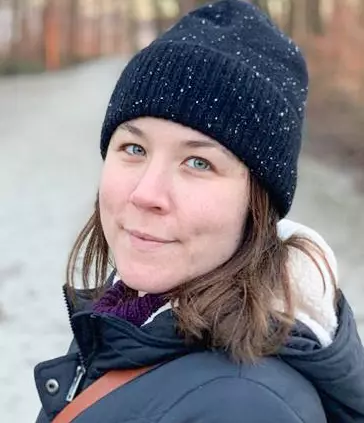}}]{Catharine Oertel}
 is an Assistant Professor at TU Delft, The Netherlands. She is co-principal investigator of the Designing Intelligence Lab (DI Lab), an effort aiming to bridge research done in computer science with industrial design engineering. Her research interest lies on understanding and modeling human interaction to build socially aware conversational agents able to engage with people in a human-like manner.
\end{IEEEbiography}

\begin{IEEEbiography}[{\includegraphics[height=1.2in,clip]{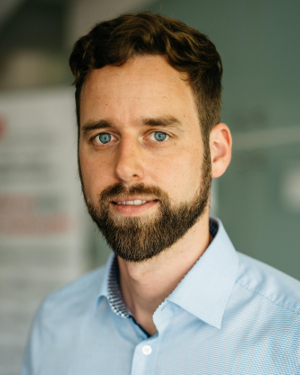}}]{Timo Gerkmann} (S’08–M’10–SM’15) is a professor for Signal Processing at the University of Hamburg, Germany. He has previously held positions at Technicolor Research $\&$ Innovation in Germany, the University of Oldenburg in Germany, KTH Royal Institute of Technology in Sweden, Ruhr-Universität Bochum in Germany, and Siemens Corporate Research in Princeton, NJ, USA. His main research interests are on statistical signal processing and machine learning for speech and audio applied to communication devices, hearing instruments, audio-visual media, and human-machine interfaces. Timo Gerkmann served as a member of the IEEE Signal Processing Society Technical Committee on Audio and Acoustic Signal Processing (2018-2023), as an Associate Editor (2019-2022) and since 2022 serves as a Senior Area Editor of the IEEE/ACM Transactions on Audio, Speech and Language Processing. He received the VDE ITG award 2022.
\end{IEEEbiography}

\begin{IEEEbiography}[{\includegraphics[height=1.1in,clip]{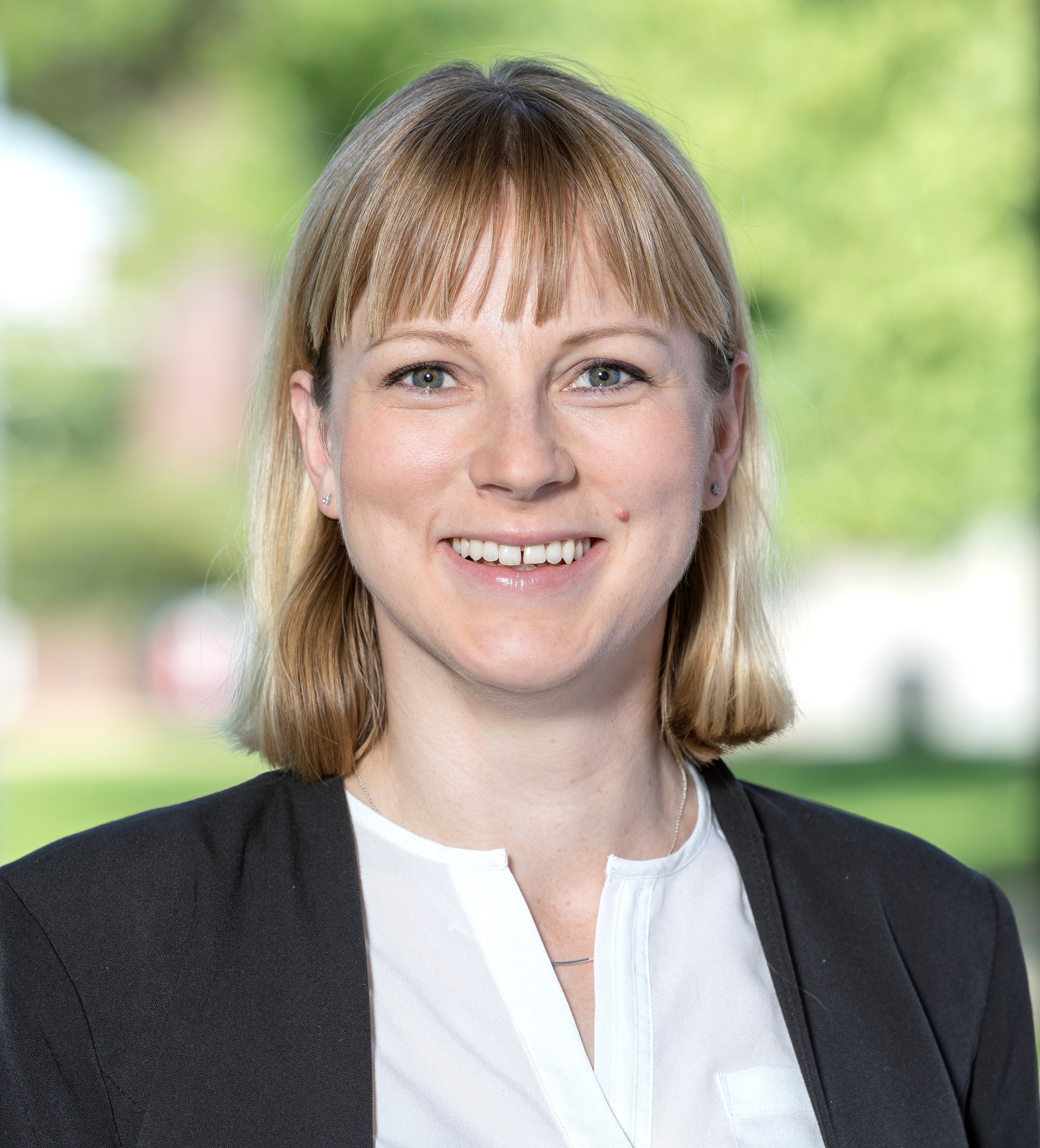}}]{Nale Lehmann-Willenbrock}
studied Psychology at the University of Goettingen and University of California, Irvine. She holds a PhD in Psychology from Technische Universität Braunschweig (2012). After several years working as an assistant professor at Vrije Universiteit Amsterdam and Associate Professor at the University of Amsterdam, she joined Universität Hamburg in 2018 as a full professor and chair of Industrial/Organizational Psychology, where she also directs the Center for Better Work. Since 2023, she also serves as Vice Dean for Research and Transfer at the Faculty of Psychology and Human Movement Science, University of Hamburg. She studies dynamic social interaction patterns in groups and teams, interpersonal processes among leaders and followers, and meetings as a core interaction site in organizations. Her research program blends organizational psychology, management, communication, and social signal processing. She currently serves as Associate Editor for the Journal of Business and Psychology.  
\end{IEEEbiography}





\end{document}


\maketitle

\section{Section 1}
Test cite \cite{avec16}.

\newpage
\bibliographystyle{IEEEtran}
\bibliography{references}
\printbibliography